%% file: Jacksonetal_2015_RLO_post-withdrawal_re-resubmission.tex
\documentclass{aastex}
\usepackage{graphicx}
\usepackage{amssymb}
\usepackage{epstopdf}
\usepackage{mathrsfs}
\usepackage{courier}
\usepackage{amsmath}

\newcommand{\be}{\begin{eqnarray}}
\newcommand{\ee}{\end{eqnarray}}
\renewcommand{\vec}[1]{\mbox{\boldmath $\displaystyle #1$}}

\newcommand{\rv}{r_{\rm v}}

\newcommand{\kepler}{{\it Kepler}}
\newcommand{\catalogretrievaldate}{2016 Dec 10}

\slugcomment{Submitted to ApJ 2016 Jul, Accepted 2016 Dec 13}
\shorttitle{Roche-lobe overflow}
\shortauthors{Jackson et al.}

\title{A New Model of Roche-lobe Overflow for Short-Period Gaseous Planets and Binary Stars}

\author{Brian Jackson}
\affil{Department of Physics, Boise State University, 1910 University Drive, Boise ID 83725-1570}
\email{bjackson@boisestate.edu}

\author{Phil Arras}
\affil{Department of Astronomy, University of Virginia, P.O. Box 400325, Charlottesville, VA 22904-4325, USA}

\author{Kaloyan Penev}
\affil{Department of Astrophysical Sciences, 4 Ivy Lane, Peyton Hall, Princeton University, Princeton, NJ 08544, USA} 

\author{Sarah Peacock}
\affil{University of Arizona, Lunar and Planetary Laboratory, 1629 E University Blvd, Tucson, AZ 85721-0092}

\author{Pablo Marchant}
\affil{Argelander Institut f{\"u}r Astronomie, Universit{\"a}t Bonn, Auf dem H{\"u}gel 71, 53121 Bonn, Germany}

\begin{abstract}
Some close-in gaseous exoplanets are nearly in Roche-lobe contact, and previous studies show tidal decay can drive hot Jupiters into contact during the main sequence of their host stars. Improving upon a previous model, we present a revised model for mass transfer in a semi-detached binary system that incorporates an extended atmosphere around the donor and allows for an arbitrary mass ratio. We apply this new formalism to hypothetical, confirmed, and candidate planetary systems to estimate mass loss rates and compare with models of evaporative mass loss. Overflow may be significant for hot Neptunes out to periods of $\sim$ 2 days, while for hot Jupiters, it may only be important inward of 0.5 days. We find that CoRoT-24 b may be losing mass at a rate of more than an Earth mass in a Gyr. The hot Jupiter WASP-12 b may lose an Earth mass in a Myr, while the putative planet orbiting a T-Tauri star PTFO8-8695 might shed its atmosphere in a few Myrs. We point out that the orbital expansion that can accompany mass transfer may be less effective than previously considered because the gas accreted by the host star removes some of the system's angular momentum from the orbit, but simple scaling arguments suggest that the Roche-lobe overflow might remain stable. Consequently, the recently discovered small planets in ultra-short-periods ($<$ 1 day) may not be the remnants of hot Jupiters/Neptunes. The new model presented here has been incorporated into Modules for Experiments in Stellar Astrophysics (MESA).

\end{abstract}


\begin{document}
 
\section{Introduction}

The origin and evolution of gaseous exoplanets with orbital periods of just a few days have challenged planetary astronomers for decades since the discovery of the first exoplanet around a Sun-like star, 51 Peg b \citep{1995Natur.378..355M}. In the decades following 51 Peg b's discovery, a bewildering variety of planets and planetary candidates have been found very close to their host stars, from hot Jupiters like 51 Peg b to hot Neptunes like GJ 1214 b \citep{2009Natur.462..891C} to hot Earths like Kepler-78 b \citep{2013ApJ...774...54S}. Early studies suggested these short-period planets formed in orbits $\gtrsim$ 1 AU and arrived near their current orbits through gas disk migration \citep{2005astro.ph..7492A} or dynamical excitation followed by tidal circularization \citep{2010fee..book..223M}. However, which and whether one of these two broad categories dominated the planets' origins remain unclear. \citet{2013MNRAS.431.3444C} considered the alternative of \emph{in-situ} formation for rocky and Neptune-like planets in periods $P < 100$ days, but \citet{2014MNRAS.440L..11R} and \citet{2014ApJ...795L..15S} have suggested such formation scenarios require unphysical or unlikely properties for the maternal protoplanetary disks. More recently, \citet{2015arXiv151109157B} revisited the idea of \emph{in-situ} formation of hot Jupiters and suggested inward migration of planetary embryos embedded in the gas disk could provide the seeds for their formation.

\begin{figure}
\includegraphics[width=\textwidth]{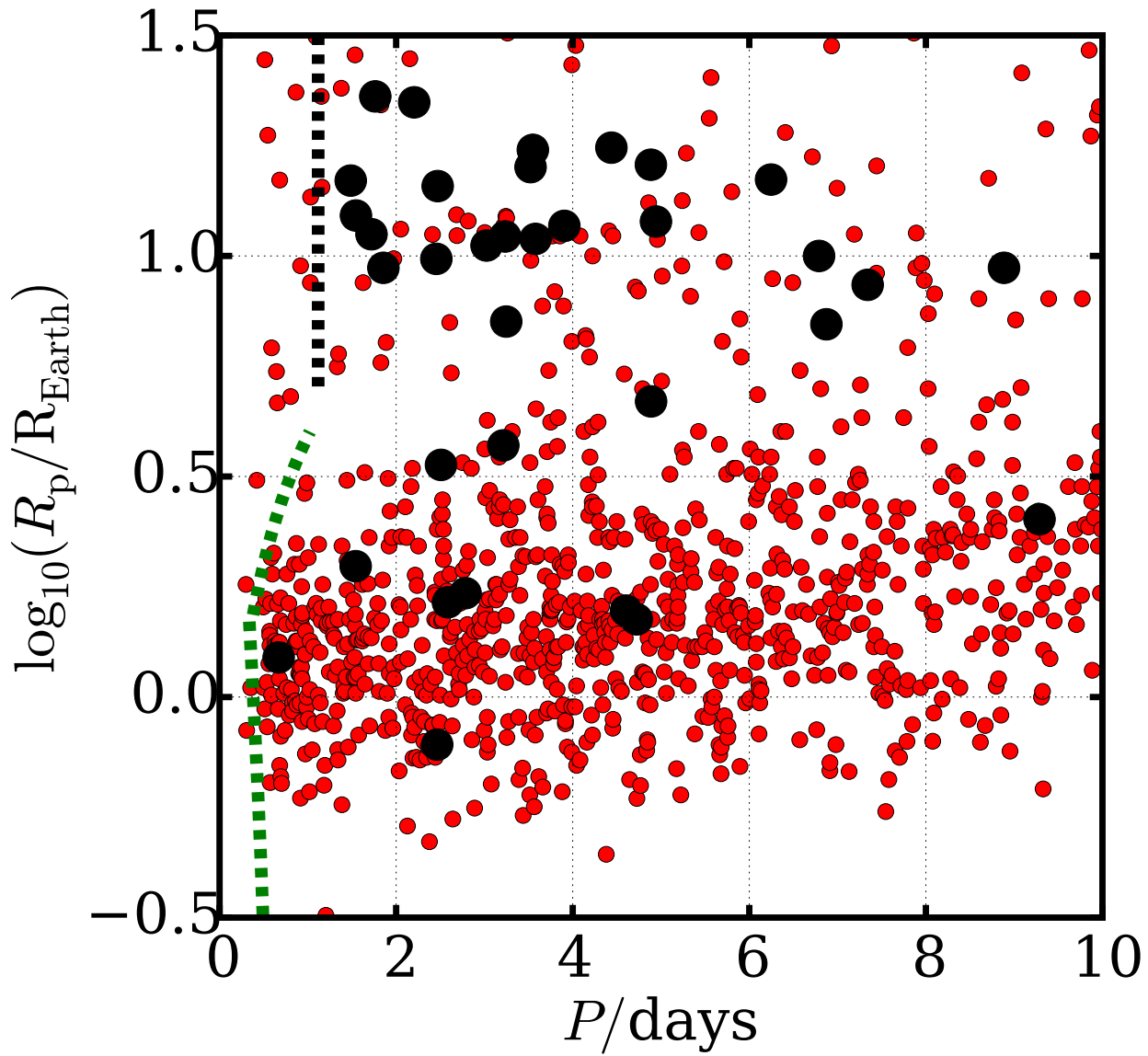}
\caption{Planetary radii $R_{\rm p}$ and orbital periods $P$ for many short-period \kepler\ confirmed planets (black circles) and candidates (red circles). The dashed black curve shows the Roche limit for a planet with a density equal to WASP-19 b's \citep{2010ApJ...708..224H}, $\rho_{\rm p} = 0.68$ g\ cm$^{-3}$, while the dashed green curve shows the Roche limit for planets obeying the relations from \citet{2014ApJ...783L...6W}.}
\label{fig:Rp_vs_P}
\end{figure}

The \kepler\ mission has been especially successful at finding such short-period planets, and Figure \ref{fig:Rp_vs_P} shows a distribution of planetary radii $R_\mathrm{p}$ and periods $P$ for \kepler\ planets and planetary candidates with $P < 10$ days for systems with only one transiting object found (data retrieved from \anchor{http://exoplanetarchive.ipac.caltech.edu/}{NASA's Exoplanet Archive} on 2014 Sep 24, and all codes and figures from this paper are available at \url{http://www.astrojack.com/research}). In the figure, the dashed black and green curves illustrate estimates for the Roche limit, the orbital period at which a planetary companion would, in principle, begin losing mass as the result of its host star's tidal gravity through the process of Roche-lobe overflow \citep[e.g.][]{1971ARA&A...9..183P}. Based on \citet{2013ApJ...773L..15R}, we estimate this period $P_\mathrm{RL}$ as 
\begin{equation}
P_\mathrm{RL} = 9.6\ \mathrm{hrs} \left(\frac{\rho_\mathrm{p}}{1\ \mathrm{g\ cm^{-3}}}\right)^{-1/2},
\label{eqn:P_L}
\end{equation}
where $\rho_\mathrm{p}$ is the planetary density. To convert $R_\mathrm{p}$ to $P_{\rm RL}$, we require $\rho_{\rm p}$, and we use the relations given by \citet{2014ApJ...783L...6W} for $R_\mathrm{p} < 4$ Earth radii $R_{\rm Earth}$:
\begin{align}
\rho_{\rm p} /\left( \mathrm{g\ cm^{-3}} \right) = 
\begin{cases} 
2.43 + 3.39 \times \left( R_{\rm p}/R_{\rm Earth} \right) & R_{\rm p} < 1.5\ R_{\rm Earth} \\ 
14.3 \times \left( R_{\rm p}/R_{\rm Earth} \right)^{-2.07} & 1.5\ R_{\rm Earth} \le R_{\rm p} \le 4\ R_{\rm Earth}
\end{cases}
\end{align}
For larger planets, we simply assume $\rho_\mathrm{p} = 0.68\ \mathrm{g\ cm^{-3}}$, consistent with the density of WASP-19 b \citep{2010ApJ...708..224H}, a typical hot Jupiter. 

From Figure \ref{fig:Rp_vs_P}, we see that many of the shortest period confirmed and candidate planets are nearly at or even interior to our simple Roche limit estimate, suggesting many planets are near Roche-lobe overflow, consistent with other evaluations. For instance, shortly after its discovery, \citet{2010Natur.463.1054L} suggested WASP-12 b, with $P \approx 1$ day and an equilibrium temperature of nearly 3,000 K, is in the process of Roche-lobe overflow, losing mass at a rate $\sim 10^{-7}$ Jupiter masses ($M_\mathrm{Jup}$) per yr. The presence of many gas giants in similar orbits shown in Figure \ref{fig:Rp_vs_P} suggests overflow may be common among exoplanets.

Even if a planet is not currently close enough to its host star to overflow, tidal interactions can drive orbital decay and endanger the planet anyway. \citet{1996ApJ...470.1187R} pointed out that the first exoplanet found around a Sun-like star, 51 Peg b, was formally unstable against tidal decay. More recently, based on the work of \citet{1973ApJ...180..307C}, \citet{2009ApJ...692L...9L} showed that the majority of short-period gas giants are formally unstable against tidal decay, and observations point to ongoing orbital decay and possible disruption of close-in exoplanets. \citet{2009ApJ...698.1357J} showed the distribution of semi-major axes and ages for planet-hosting stars was qualitatively consistent with disruption and removal of close-in exoplanets. \citet{2014A&A...565L...1P} found, comparing members of binary star systems, the stars with hot Jupiters and deep convective zones (where tidal dissipation is thought to be efficient -- \citealp{2012ApJ...757...18A}) exhibited faster spin rates than their planetless companions, consistent with tidal spin-up. Estimating spin periods for \kepler\ target stars, \citet{2013ApJ...775L..11M} found rapidly rotating stars are less likely to host close-in planets, qualitatively consistent with the removal of planets and spin-up of stars by tidal decay. Thus, tidal decay and disruption likely shapes the population of close-in exoplanets. 

Several studies have investigated these processes in detail. One of the earliest on the topic, \citet{1998ApJ...500..428T} considered the combined effects of tidal decay, planetary disruption, and gas disk migration on the then newly discovered population of hot Jupiters. That study predicted some planets would experience overflow or be accreted by their host stars during a planetary system's adolescence. \citet{2003ApJ...588..509G} included tidal heating of a gas giant from a non-zero orbital eccentricity and showed heating from even fairly small eccentricities can inflate a planet and drive it into Roche-lobe overflow. \citet{2012MNRAS.425.2778M} suggested disruption and accretion of planetary matter could produce optical and x-ray transient events for planet-hosting stars. 

\citet{2014ApJ...793L...3V} studied planetary Roche-lobe overflow in light of \kepler\ discoveries of small, ultra-short-period planets \citep{2013ApJ...779..165J, 2014ApJ...787...47S, 2016arXiv160306488A}. That study higlighted the fact that exchange of momentum between the escaping atmosphere and an overflowing planet can drive orbital expansion, and so a gas giant that begins overflowing at $P \sim 0.3$ day can end up as a denuded rocky core with $P$ between a few hours and a few days, possibly accounting for the small, short-period \kepler\ planets (although even gas giants formed via core accretion may not have rocky cores to leave behind -- \citealp{2012ApJ...745...54W}). As follow-on studies, \citet{2015ApJ...813..101V} found that the final $P$-value depends sensitively on the planet's internal response to mass loss, and \citet{2016arXiv160300392J} found that the exact orbital evolution depends most sensitively on the mass of the gas giant's core.

The orbital evolution of an overflowing planet depends on the rate of mass loss $\dot{M}$, and, in turn, $\dot{M}$ depends on the planet's proximity to its star. Thus, the accuracy of predictions for the fates of hot Jupiters and the final orbits of their remnants depends on the accuracy of the loss model and the mass-radius relationship used. Overflow models often approximate the donor as having a discrete outer boundary at the photosphere. Consequently, the mass loss rate is assumed to be zero until the pressure contour corresponding to the planet's photosphere exactly coincides with the Roche lobe gravitational contour, and then mass loss becomes non-zero at the instant of Roche lobe contact. In reality, though, the atmospheres of stars and planets taper off into space, and so overflow gradually increases from very small values as a donor's Roche lobe closes in on the donor. For close-in planets, this effect is non-negligible since their upper atmospheres can be very hot and distended. Moreover, the question of whether the mass loss is stable or runs away depends sensitively on how the loss rate evolves with distance between the donor and accretor.

Motivated by these considerations, we present a new model for Roche-lobe overflow. The model involves a 1-D treatment of the hydrodynamics of outflow and can be applied to a very wide range of mass ratios for the donor and accretor, including binary stars and short-period planets. In Section \ref{sec:model}, we formulate the model and compare it to the model developed by \citet{1988A&A...202...93R} upon which it is based. In Section \ref{sec:results}, we apply the model to hypothetical, confirmed, and candidate planetary systems. Finally, in Section \ref{sec:conclusions}, we discuss implications of our predictions and future work.

As of version 7499, the model presented here has been incorporated into Modules for Experiments in Stellar Astrophysics, MESA \citep{2011ApJS..192....3P, 2013ApJS..208....4P}. It can be activated by setting \texttt{mdot\_scheme = `Arras'} in a project inlist. See \url{http://mesa.sourceforge.net/} for details. It is worth noting that one of the previous overflow schemes provided by MESA, \texttt{`Ritter'}, had some bugs in the code, which have been corrected as of this version. The \texttt{`Ritter'} model also did not actually apply outside of a narrow range of mass ratios, particularly for typical planet-star mass ratios, as we discuss below.

\section{The Mass Transfer Model}
\label{sec:model}

\subsection{Model Formulation}
The Roche-lobe overflow model described here resembles closely models presented in previous work, primarily that of \citet{1988A&A...202...93R} but also that of \citet{1972AcA....22...73P} and work from an unpublished thesis \citep{Jedrzejec1969}. We consider isothermal mass loss through the L1 Lagrange point from a donor (either star or planet with an extended atmosphere). As highlighted by \citet{1988A&A...202...93R}, the atmosphere of the donor does not terminate at a hard boundary but rather has a finite extension above the photosphere. Thus, even if the volume of the donor's photosphere does not extend to its Roche lobe, the density of gas near the Roche lobe may be non-negligible, allowing overflow to occur. The mass loss rate $\dot{M}_{\rm d}$ from the donor is given by
\begin{equation*}
\dot{M}_{\rm d} = -  \int ({\rm density\ at\ L1}) \times ({\rm velocity\ at\ L1}) \ d({\rm area}),
\end{equation*}
where the integral is taken over the area of the outflow nozzle around L1. In the following section, we will show the final result is given by
\be
\dot{M}_{\rm d} & = & - 
\left[ \frac{1}{\sqrt{e}} \rho_{\rm ph}\ e^{-\left( \Phi_{\rm L1}-\Phi_{\rm ph} \right)/v^2_{\rm th} } \right] 
\times v_{\rm th}
\times
\left[ \frac{2\pi v_{\rm th}^2/\Omega^2}{ \sqrt{A(A-1)}} \right],
\label{eq:mdot}
\ee
where $\rho_{\rm ph}$ and $\Phi_{\rm ph}$ are the density and effective potential at the photosphere of the donor, $\Phi_{\rm L1}$ is the potential at the L1 point, and $\upsilon_{\rm th}$ is the isothermal sound speed ($\upsilon_{\rm th} = \sqrt{k_{\rm B} T/\mu}$, with $k_{\rm B}$ Boltzmann's constant, $T$ the atmospheric temperature, and $\mu$ the average mass of atmospheric particles). $A$ is a dimensionless coefficient that depends on the ratio $q$ of the donor mass $M_{\rm d}$ to the accretor mass $M_{\rm a}$. $\Omega^2 = \left( G(M_{\rm d}+M_{\rm a}) \right)/a^3$ is the orbital frequency squared, with $G$ the gravitational constant and $a$ the orbital semi-major axis. Note that our definition for the mass ratio $q = M_{\rm d}/M_{\rm a}$ is not the same as that used by \citet{1988A&A...202...93R}, where it is defined as $M_{\rm a}/M_{\rm d}$.

First, we solve for the area of the outflow nozzle near L1. Following \citet{1975ApJ...198..383L}, consider a coordinate system revolving with the binary and centered on the donor with the $x$-axis pointing from the donor to the accretor (which are separated by a fixed orbital distance $a$) and the $z$-axis pointing parallel to the orbital angular momentum. The $x$-position of the center of mass $x_{\rm cm}$ for the binary system is
\be
x_{\rm cm} & = & a \left( \frac{M_{\rm a}}{M_{\rm a}+M_{\rm d}} \right), 
\ee
and the effective potential (gravitational + centrifugal) field $\Phi$ is approximately
\be
\Phi = - \frac{GM_{\rm d}}{|\vec{r}|} - \frac{GM_{\rm a}}{| a \hat{x} - \vec{r} |} - \frac{1}{2} \frac{G(M_{\rm d}+M_{\rm a})}{a^3} 
\left[ \left(x - x_{\rm cm}\right)^2 + y^2 \right],
\label{eq:phi_full}
\ee
where $\vec{r}$ is the location at which to evaluate $\Phi$ and $\hat{x}$ is a unit vector pointing along $x$. 

Expanded to 2nd order about the L1 point at $ x_{\rm L1} $,
\be
\Phi & \approx & \Phi_{\rm L1} + \frac{1}{2}\Omega^2 \left[ -(1+2A)(x-x_{L1})^2 + (A-1)y^2 + Az^2 \right],
\label{eq:phi}
\ee
and 
\be
A & = & \frac{a^3}{M_{\rm d}+M_{\rm a}} \left[ \frac{M_{\rm d}}{|x_{\rm L1}|^3} + \frac{M_{\rm a}}{ |x_{\rm L1} - a|^3} \right].
\label{eq:A_exact}
\ee

\noindent Along the y- and z-directions near L1, the escaping, (assumed) isothermal gas is nearly in hydrostatic balance \citep{1975ApJ...198..383L}, and so the fluid density falls off exponentially in both directions, i.e. $\rho \propto e^{-\left( \Phi_{\rm L1} - \Phi \right)/v_{\rm th}^2}$. Thus, most of the mass escapes within a scale height of L1 (with scale heights $\Delta y$ and $\Delta z$ in the y- and z-directions respectively), i.e. through an elliptical nozzle centered on L1 with an area $\pi \Delta y \Delta z$, and 
\be
\Delta y & = & \frac{\sqrt{2} v_{\rm th}}{\sqrt{A-1} \Omega},
\\
\Delta z & = & \frac{\sqrt{2} v_{\rm th}}{\sqrt{A} \Omega}.
\ee

\begin{figure}[h]
\includegraphics[width=\textwidth]{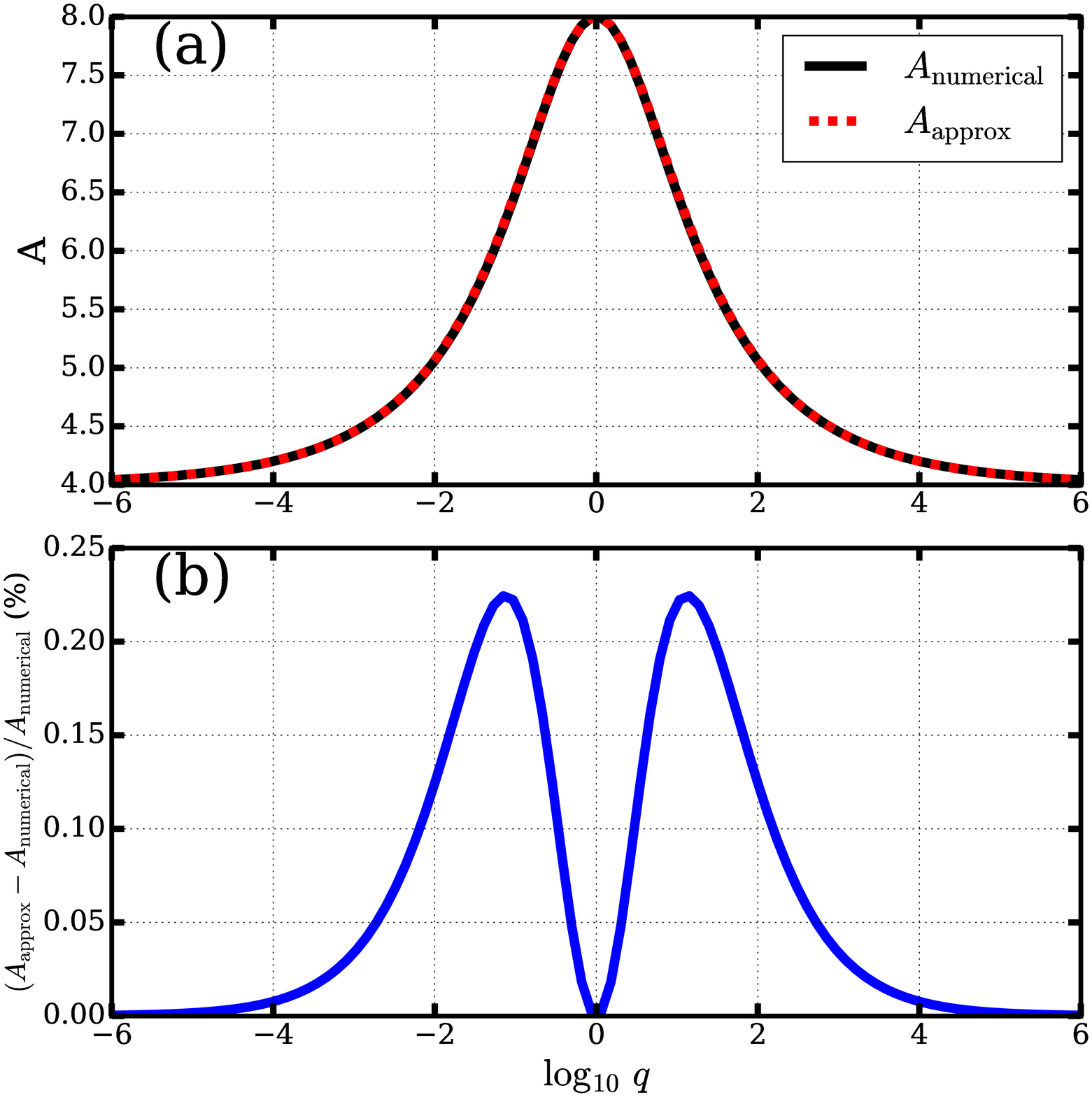}
\caption{The dimension quantity $A$ from Equation \ref{eq:A_exact}. (a) The approximate root for Equation \ref{eq:A_approx}, $A_{\rm approx}$, compared to a more accurate solution found by numerically solving for the roots, $A_{\rm numerical}$, both as functions of the ratio $q$ of the donor's $M_{\rm d}$ to the accretor's $M_{\rm a}$ mass. (b) The difference between the two solutions, scaled by the more accurate one.}
\label{fig:coeff_A}
\end{figure}

\noindent As illustrated in Figure \ref{fig:coeff_A}, the coefficient $A$ only varies by a factor of 2 for the entire range of $q$. For an equal mass binary, $M_{\rm d}=M_{\rm a}$ and $x_{\rm L1}=a/2$, giving $A=8$. For $M_{\rm d} \ll M_{\rm a}$, $x_{\rm L1} \simeq a (M_{\rm d}/3M_{\rm a})^{1/3}$, giving $A \simeq 4 + 3(M_{\rm d}/3M_{\rm a})^{1/3} \simeq 4$. The following expression fits $A(q)$ to an accuracy of 0.3\% for the whole range of $q$ (Figure \ref{fig:coeff_A}): 
\be
A(q) & = & 4 + \frac{b_1}{b_2 + q^{1/3} + q^{-1/3}}.
\label{eq:A_approx}
\ee
with $b_1=2 \cdot 3^{2/3}=4.16$ and $b_2=b_1/4-2=-0.96$. Since the function is symmetric about $M_{\rm d}=M_{\rm a}$, either $M_{\rm d}/M_{\rm a}$ or $M_{\rm a}/M_{\rm d}$ can be used as an argument.

Turning to the first term in Equation \ref{eq:mdot}, the density near the L1 point can be expressed in terms of the density at the donor's photosphere. As \citet{1988A&A...202...93R}, consider the Bernoulli integral for the streamline of isothermal fluid passing from the photosphere to L1:
\be
\left( \frac{1}{2} v^2 + v_{\rm th}^2 \ln \rho + \Phi  \right)_{\rm ph} & = & \left( \frac{1}{2}v^2  + v_{\rm th}^2 \ln \rho + \Phi \right)_{\rm L1}, 
\ee
where $\upsilon$ is the fluid velocity and $\rho$ the fluid density. At the photosphere, $v \ll v_{\rm th}$, while near L1, the gas is assumed to reach the isothermal sound speed, $v \simeq v_{\rm th} $, giving
\be
v_{\rm th}^2 \ln \left( \frac{\rho_{\rm L1}}{\rho_{\rm ph}} \right) & = & - \frac{1}{2} v_{\rm th}^2 + \Phi_{\rm ph} - \Phi_{\rm L1}
\ee
or
\be
\rho_{\rm L1} & = & \frac{1}{\sqrt{e}} \rho_{\rm ph}\ e^{-(\Phi_{\rm L1}-\Phi_{\rm ph})/v_{\rm th}^2}.
\label{eq:rhoL1}
\ee

To evaluate this density, we need to calculate the potential difference between the L1 point and the photosphere, $\Phi_{\rm L1}-\Phi_{\rm ph}$. In the Roche model, isopotential surfaces enclosing the donor are not strictly spherical, although some previous investigators have expanded the potential difference $\Phi_{\rm L1}-\Phi_{\rm ph}$ in terms of the distance from L1 to the donor's photosphere. A more accurate approach is to find a sphere with a radius $\rv$ and volume $V(\Phi)$ equivalent to that enclosed by an equipotential $\Phi$, i.e.\ $V(\Phi) \equiv 4\pi r_{\rm v}^3/3$, so that $r_{\rm v} = r_{\rm v}(\Phi)$. Given a volume-equivalent radius for the donor photosphere $r_{\rm ph}$ and its Roche lobe $r_{\rm R}$, we can compute the potential difference using a relation between $\Phi$ and $\rv$, as needed by Equation \ref{eq:rhoL1} to compute the density at the L1 point. Note that \citet{1983ApJ...268..368E} gives a good approximation for $r_{\rm R} = \dfrac{0.49 q^{2/3}}{0.6 q^{2/3} + \ln\left( 1 + q^{1/3} \right)}\ a$. \citet{1983ApJ...268..368E} reports this expression is accurate to 1\% for all $q$.

In Appendix \ref{apx:potential}, we derive the volume-averaged potential $\Phi(r_{\rm v})$, given by 
\be
\Phi(\rv) = &-& \left(  \frac{GM_{\rm a}}{a} + \frac{ GM_{\rm a}^2 }{2a(M_{\rm d}+M_{\rm a})} \right)
\nonumber \\ & - & 
 \frac{GM_{\rm d}}{\rv} \left[ 1 + \frac{1}{3} \left( \frac{M_{\rm d}+M_{\rm a}}{M_{\rm d}} \right) \left( \frac{\rv}{a} \right)^3
 \right. \nonumber \\ & + & \left.
 \frac{4}{45} \left( \frac{(M_{\rm d}+M_{\rm a})^2 + 9M_{\rm a}^2 + 3M_{\rm a}(M_{\rm d}+M_{\rm a})}{M_{\rm d}^2}\right) \left( \frac{\rv}{a} \right)^6
\right].
\label{eq:phi_vs_rv}
\ee

\begin{figure}
\includegraphics[width=\textwidth]{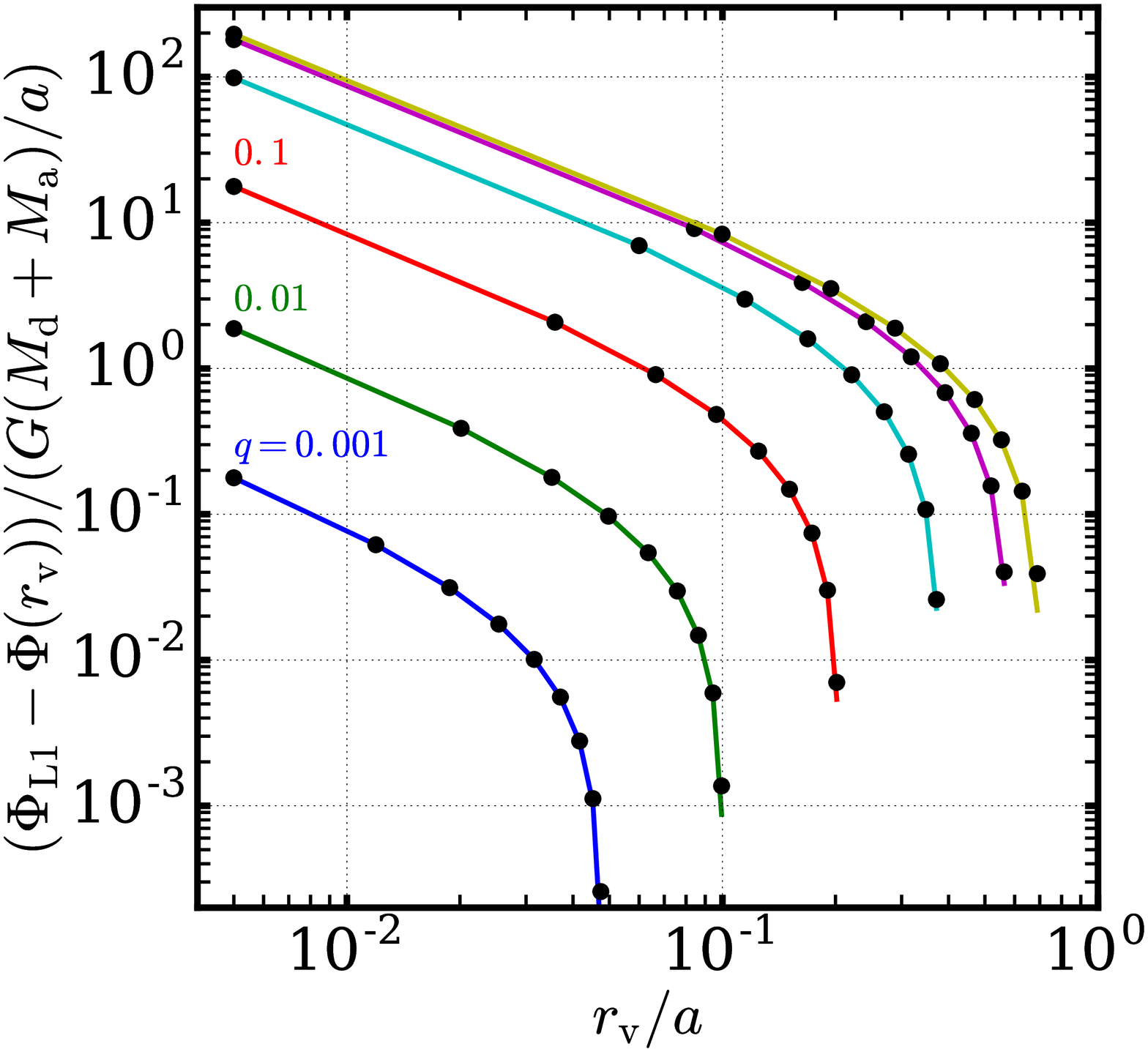}
\caption{Relationship between the potential surfaces and the radius of a sphere with the same volume $\rv$ for a range of mass ratios $q$, from 0.001 to 100. The solid curves represent the approximate solution, Equation \ref{eq:phi_vs_rv}, and the black dots numerical solutions.}
\label{fig:potential_vs_volume_radius}
\end{figure}

The first term in parentheses on the right-hand side of the equation is a constant and can be ignored. The term in square brackets contains an expansion in the parameter $\rv/a$ that converges inside the Roche lobe but not outside. The first term in the square brackets corresponds to gravity of the donor as a point mass. The next term represents rotational distortion of the donor (assumed to rotate with the orbital frequency), while the final term represents distortion from rotation, tides, and a cross term. Plugging in $r_{\rm R}$ for $r_{\rm v}$ gives $\Phi_{\rm L1}$. Figure \ref{fig:potential_vs_volume_radius} compares this approximate solution for $\Phi$ to a numerical solution. For the latter, we numerically solved for the L1 Lagrange point and directly evaluated the Roche potential there. The solutions diverge as $\rv$ approaches $r_{\rm R}$ (as our approximations break down) but agree to better than 2\% for the range of parameters shown. (Equation \ref{eq:rv3_in_terms_of_potential_expansion} in Appendix \ref{apx:potential} provides a means to convert the donor's photospheric $R_{\rm p}$ radius to the volume-averaged radius $r_{\rm ph}$, which in most cases agree to a few percent or better.)

To estimate the error in the formulation of potentials presented here, we can compare the difference in potential between a donor's photosphere and the Roche lobe for our approach to a numerical estimate of the volume-average potentials (estimated to within $1\times10^{-8}$). We find the two approaches agree to a few percent for the range of parameters explored here. Since the potential difference appears in the exponential (Equation \ref{eq:mdot}), this error amounts to about a factor of two in the final mass loss rate for planets, but, of course, the exact error depends on the other parameters entering into the equation. In particular, the uncertainties involved in choosing the atmospheric temperature (as discussed below in Section \ref{sec:results}) dominate the uncertainty in the mass loss rate estimate.

Our approach parallels some of the results from \citet{1984ApJS...55..551M}, in which the Roche model is integrated to give volume radii and surface areas, among other results. Equation 14 from that article provides the approximate normalized Roche potential but has two small typos and should read
\be
C \approx \dfrac{2}{1 + q}\dfrac{1}{\overline{r}_1} + \dfrac{2 q}{1 + q} + \left( \dfrac{q}{1 + q} \right)^2 + \dfrac{\overline{r}_1^2}{3},
\label{eq:Mochnacki_eq}
\ee
where $q$ is the mass ratio for the binary system and $\overline{r}_1$ is the volume-averaged Roche lobe radius normalized to the orbital semi-major axis. Our Equation \ref{eq:phi_vs_rv} goes one order further than Equation \ref{eq:Mochnacki_eq}.

\subsection{Comparison to the work of \citet{1988A&A...202...93R}}
The model presented here is motivated by that described by \citet{1988A&A...202...93R}, but there are some important differences and improvements over that previous model, which we explore in this section.

To begin with, \citet{1988A&A...202...93R} expands the potential difference about the L1 point to the donor as follows:
\be
\Phi_{\rm L1}-\Phi_{\rm ph} & \approx & \frac{d\Phi}{d\rv} \rfloor_{\rm L1}  \left( r_{\rm R} - r_{\rm ph} \right),
\ee
where $r_{\rm ph}$ is the donor's photospheric radius. This Taylor expansion may underestimate the potential difference when the donor is well inside its Roche lobe radius and hence overestimate the mass loss rate. \citet{1988A&A...202...93R} goes on to define the relationship
\be
\frac{d\Phi}{d\rv} \rfloor_{\rm L1} & =& \frac{GM_{\rm d}}{r_{\rm R}^2} \gamma(q).
\ee
(It is important to note that the dimensionless function $\gamma(q)$ from \citet{1988A&A...202...93R} depends on the ratio $1/q = M_{\rm a}/M_{\rm d}$.) 
These relations cast the exponential argument in the expression for the density near L1 as
\be
\frac{ \Phi_{\rm L1}-\Phi_{\rm ph}}{v_{\rm th}^2} & \equiv & \gamma  \left( \frac{ r_{\rm R} - r_{\rm ph} }{ H_{p,0} } \right)
\ee
where 
\be
\frac{\gamma}{H_{p,0}} &  = & \frac{1}{v_{\rm th}^2} \  \frac{d\Phi}{d\rv} \rfloor_{\rm L1}.
\label{eqn:modified_scale_height}
\ee
Here $H_{p,0}$ is planet's atmospheric scale height in the absence of tidal gravity, and $\gamma$ is the tidal correction.

\begin{figure}
\includegraphics[width=\textwidth]{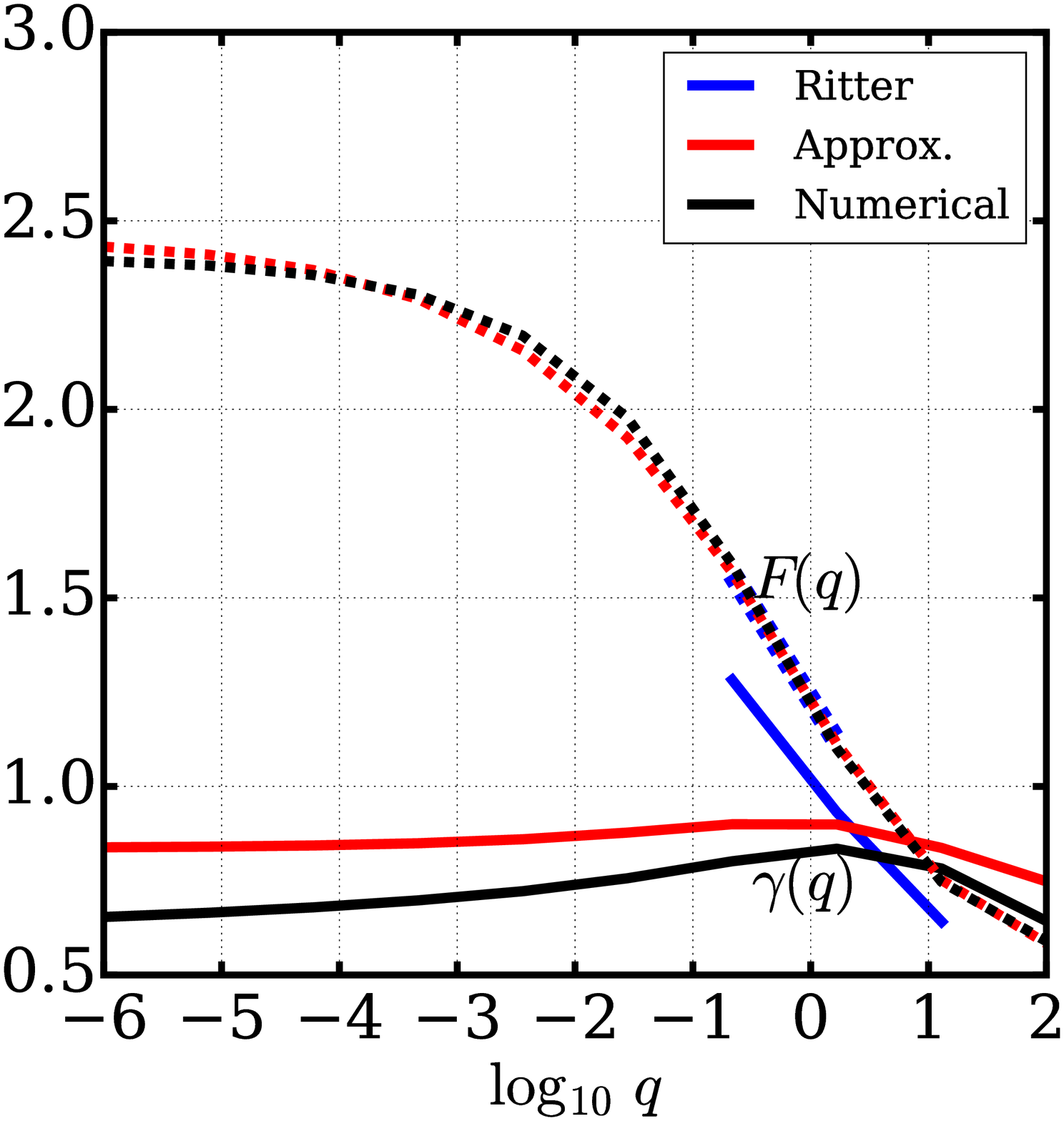}
\caption{The solid lines illustrate $\gamma(q)$ and the dashed $F(q)$. As described in the text, $\gamma(q)$ is the correction to the atmospheric scale height due to the tidal gravity (used in Equation \ref{eqn:modified_scale_height}), and $F(q)$ is the normalized cross-sectional area of the outflow nozzle near L1 (used in Equation \ref{eqn:ritter_nozzle_area}). The blue lines use the formulae from \citet{1988A&A...202...93R}, the red the corresponding forms based on the formulation proposed here, and the black direct numerical estimates. The dashed blue curve for Ritter's $F(q)$ lies beneath the other two dashed curves.}
\label{fig:compare_gamma-F}
\end{figure}

Equation 7 from \citet{1988A&A...202...93R} provides a numerical fit to $\gamma$:
\begin{align}
\gamma = 
\begin{cases} 
0.954 + 0.025 \ln \left( M_{\rm a}/M_{\rm d}\right) - 0.038 \left[ \ln \left( M_{\rm a}/M_{\rm d} \right) \right]^2, {\rm 0.04} \le M_{\rm a}/M_{\rm d} \le {\rm 1}\\
 0.954 + 0.039 \ln \left( M_{\rm a}/M_{\rm d} \right) + 0.114 \left[ \ln \left( M_{\rm a}/M_{\rm d} \right) \right]^2, {\rm 1} \le M_{\rm a}/M_{\rm d} \le {\rm 20}
\end{cases}
\label{eq:ritter_gamma}
\end{align}
Although \citet{1988A&A...202...93R} reports these expressions fit a numerical integration result to better than 1\% for the range of $M_{\rm a}/M_{\rm d}$ given, they are inadequate for disruption of planets, where $M_{\rm a}/M_{\rm d} \sim $ 1000. We can re-cast our Equation \ref{eq:phi_vs_rv} in similar terms for comparison, and Figure \ref{fig:compare_gamma-F} shows how our formulation for $\gamma$ compares to that from \citet{1988A&A...202...93R} and to a numerical solution. Our formulation agrees to within 30\% of the numerical solution, while that from \citet{1988A&A...202...93R} diverges considerably. In any case, the new model we present here does not use the $\gamma$ approach but the fit in Equation \ref{eq:phi_vs_rv}.

For calculating the nozzle area through which escape occurs, \citet{1988A&A...202...93R} defines the $F(q)$ function: 
\be
\frac{2\pi (v_{\rm th}/\Omega)^2}{ \sqrt{ A(A-1)} } \equiv 2\pi v_{\rm th}^2 \frac{r_{\rm R}^3}{GM_{\rm d}} F(q).
\label{eqn:ritter_nozzle_area}
\ee
\citet{1988A&A...202...93R} provides the following fit (Equation A9): 
\be
F\left( M_{\rm a}/M_{\rm d} \right) = 1.23 + 0.5 \ln\left( M_{\rm a}/M_{\rm d} \right), 0.5 \le \left( M_{\rm a}/M_{\rm d} \right) \le 10.
\label{eq:ritter_F}
\ee
Again, Figure \ref{fig:compare_gamma-F} compares this formulation to ours and to a numerical solution, and both agree with the numerical to within 2\% over the limited range of $q$ where Equation \ref{eq:ritter_F} applies. However, Equation \ref{eq:ritter_F} only applies over a limited range of $q$ and breaks down for small $q$ (a massive donor), becoming negative for $q \approx$ 288.

\begin{figure}
\includegraphics[width=\textwidth]{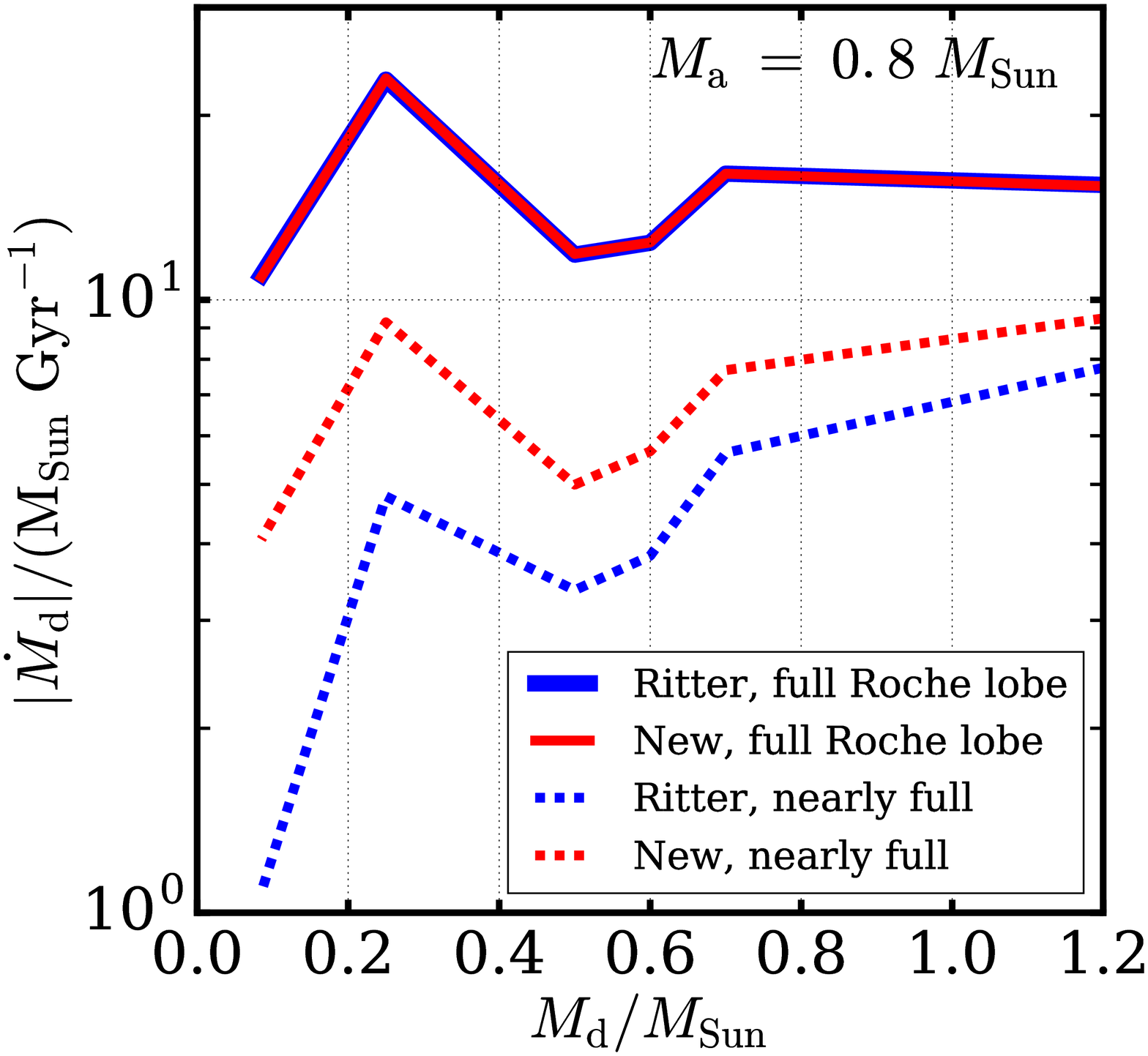}
\caption{Mass loss rates for the parameters for low-mass binary stars given in Table A1 from \citet{1988A&A...202...93R} and reproduced in Table \ref{tbl:ritter_table_a1}. The solid lines (which lie on top of each other) show loss rates for a donor filling its Roche lobe. The dashed lines show loss for a donor ``nearly filling'' its Roche lobe (i.e. $\left( r_{\rm R} - r_{\rm ph} \right)/r_{\rm R} = 0.01\%$ for the blue lines and $\left( \Phi_{\rm ph} - \Phi_{\rm R} \right)/\Phi_{\rm R} = 0.01\%$) for the red.}
\label{fig:compare_mdot}
\end{figure}

Finally, Figure \ref{fig:compare_mdot} compares the mass loss rates predicted by \citet{1988A&A...202...93R} and our new model here for parameters for low-mass binary stars given in the former study's Table A1, reproduced in Table \ref{tbl:ritter_table_a1}. Note that, in order to directly compare the two models, we require an assumed $M_{\rm a}$ to evaluate $\Phi$. We chose $M_{\rm a} = 0.8\ M_{\rm Sun}$ so $q$ stayed in the range considered by \citet{1988A&A...202...93R}. The solid lines in Figure \ref{fig:compare_mdot} show the case where the donor is filling its Roche lobe, i.e. $r_{\rm ph} - r_{\rm R} = 0$ for the model from \citet{1988A&A...202...93R} (red lines) and $\Phi_{\rm ph} - \Phi_{\rm R} = 0$ for ours (blue lines). The two models agree well, as they should since our estimates of the nozzle area agree and the $\gamma$ function doesn't factor in when the donor's photosphere fills the Roche lobe. The non-monotonic variation in loss rate is due to the combination of varying stellar mass and composition.

\input{Ritter_Table_A1.tex}

The dashed lines show the two models for nearly full Roche lobes: $\left( r_{\rm R} - r_{\rm ph} \right)/r_{\rm R} = 0.01\%$ and $\left( \Phi_{\rm ph} - \Phi_{\rm R} \right)/\Phi_{\rm R} = 0.01\%$, respectively. Disagreement between these models is considerable since the $\gamma$ function from \citet{1988A&A...202...93R} (which determines the donor's atmospheric scale height) disagrees significantly from our effective $\gamma$. It is worth noting here that if we did not consider the fact that the donor's atmosphere is extended, $\dot{M}_{\rm d}$ in Figure \ref{fig:compare_mdot} would be zero for the nearly full cases, while the figure shows significant mass loss.

\section{Application to Short-Period Gaseous Planets}
\label{sec:results}

In this section, we apply our new model to hypothetical, candidate, and confirmed short-period planetary systems. The estimates of Roche-lobe overflow calculated using the new model presented here involve several assumptions since, in many cases, all the planetary parameters required are not available.

Where unavailable from observations, we estimated a planet's photospheric temperature $T_{\rm p}$ by assuming the planet's dayside is a blackbody at radiative equilibrium with its star (i.e., we ignore redistribution of stellar insolation): $T_{\rm p} = T_* \sqrt{R_\star/(2^{1/2} a)}$, where $R_\star/T_\star$ is the stellar radius/effective temperature. To estimate photospheric density $\rho_{\rm ph}$ from a planet's mass $M_{\rm p}$ and radius $R_{\rm p}$, we first estimated the pressure scale height $H$ as $H = k_{\rm B} T_{\rm p}/(\mu g)$, where $g$ is the planet's surface gravity. Then, we calculated $\rho_{\rm ph} = \tau/\left( \kappa \sqrt{2 \pi R_{\rm p} H} \right)$, where $\tau$ is the slant optical depth along the planet's photospheric limb, which we take as 0.56 \citep[][Appendix]{2012ApJ...756..176H}. 

We make the simple approximation that the atmosphere is composed entirely of molecular hydrogen ($\mu =$ 2 amu) when $T_{\rm p} <$ 2000 K and is composed entirely of atomic hydrogen ($\mu =$ 1 amu) above that temperature. Atmospheric studies of evaporating exoplanets suggest a fairly abrupt transition from one to the other at about that temprature \citep{2004Icar..170..167Y}, but this assumption represents an important approximation that may limit the accuracy of our estimates here. We also take a simple prescription for $\kappa$, the Rosseland mean opacity: we assume it is equal to $10^{-2}\ {\rm cm^2\ g^{-1}}$, as assumed in \citet{2010Natur.463.1054L} and based on Figure 2 from \citet{2008ApJS..174..504F}. The latter study showed that, for hot Jupiter atmospheres with solar metallicity, $\kappa$ is insensitive to pressure and temperature (at least for temperatures $\sim$ 2000 K). In any case, a more complete model from MESA can provide more accurate estimates for all the relevant atmospheric parameters, but our simple scheme here provides representative loss rates. 

It is important to re-iterate that our results assume an isothermal escape. By definition, the optical depth drops off above the photosphere, and so, along the outflow, the gas temperature is more and more sensitive to the balance between radiative heating and cooling. Atmospheric mass loss is neither entirely adiabatic nor isothermal, and at these altitudes, molecular hydrogen can be disassociated and atomic hydrogen can be photoionized. \citet{2011ApJ...728..152T} showed temperatures in the upper atmosphere can greatly exceed the equilibrium temperature, reaching $\sim 10^4$ K but holding roughly constant with altitude thereafter. \citet{2009ApJ...693...23M} explored the cooling and heating of escaping atmospheres and found a complex balance between heating and cooling that depends on the details of the outflow. We have neglected these complications here, but future work may produce a model incorporating these effects but still simple enough for inclusion in an evolutionary code.

\subsection{Hypothetical Systems}

\begin{figure}
\includegraphics[width=\textwidth]{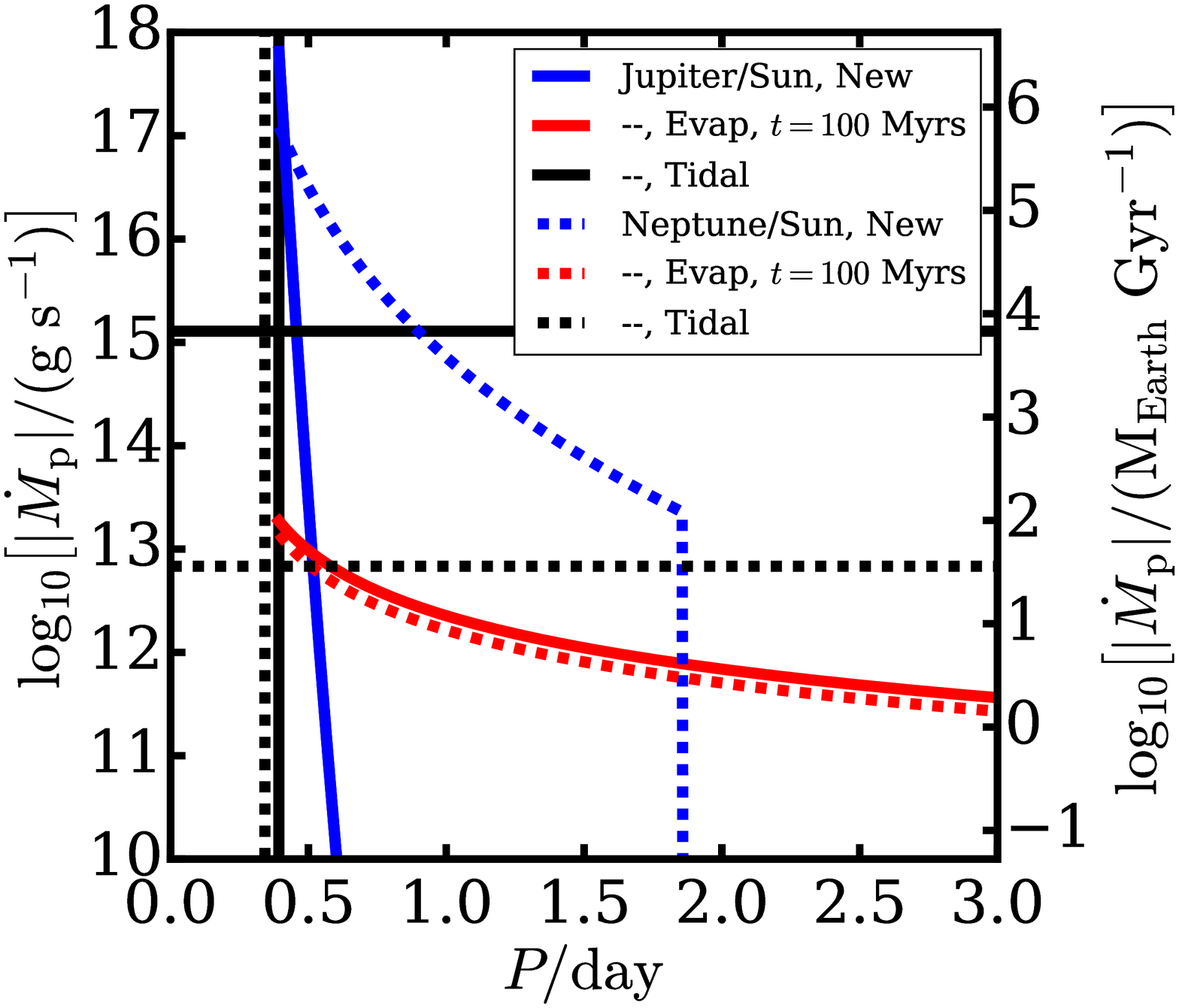}
\caption{Planetary mass loss rates. The blue lines show the rates from the Roche-lobe overflow model presented here (``New''), the red lines the evaporative mass loss (``Evap'') \citep[e.g.][]{2012ApJ...761...59L} assuming a system age of 100 Myrs, and the black lines the torque-balance model (``Tidal''). The solid lines show loss for Jupiter around the Sun, and the dashed for Neptune around the Sun. The Roche limits for the hypothetical Jupiter/Neptune are also shown as vertical solid/dashed black lines.}
\label{fig:Mdot_map}
\end{figure}

Figure \ref{fig:Mdot_map} shows predicted mass loss rates for a Jupiter-like and a Neptune-like planet with $R_\mathrm{\rm p}$ and $M_{\rm p}$ equal to Jupiter's (71492 km and 1.8987$\times10^{30}$ g, respectively)/Neptune's (24622 km and 102.4$\times10^{24}$ kg, respectively) in a short-period orbit around a Sun-like star (with  $R_\mathrm\star$ and mass $M_\star$ equal to the Sun's). We assume the star has a Sun-like effective temperature $T_\star =$ 6000 K. Our loss rate is shown with a solid (for the Jupiter) and a dashed (for the Neptune) blue line and denoted as ``New'' in the figure. 

Figure \ref{fig:Mdot_map} also shows loss due to atmospheric evaporation (``Evap'') as a solid and dashed red line and estimated as 
\begin{equation}
\dot{M}_{\rm p, evap} = \frac{\pi \epsilon R_{\rm p}^3 F_{\rm XUV}}{G M_{\rm p} K_{\rm tide}}.
\end{equation}
This equation is used \citet{2012ApJ...761...59L}, among many other studies, and $F_{\rm XUV}$ is the stellar X-ray and ultraviolet (XUV) flux evaluated at an age of 100 Myr using results from \citet{2005ApJ...622..680R}. $\epsilon$ is the fraction of XUV flux that powers mass loss (which we take as 0.1 -- \citealp{2012ApJ...761...59L}), and $K_{\rm tide}$ is an enhancement in the loss rate due to the stellar tidal gravity \citep{2007A&A...472..329E}, is unitless, and takes a value between 0 (a planet at its Roche limit) and 1 (a planet very far from its Roche limit). 

Figure \ref{fig:Mdot_map} shows that, for the hypothetical planets considered here, evaporative mass loss dominates over Roche-lobe overflow for the longest periods. Closer in, which process dominates depends sensitively on the planet-star proximity and the planet's density. The ``New'' loss rate for the Neptune dominates inward of about 2 day, where $T_{\rm p}$ exceeds 2000 K and so we assume an atomic hydrogen atmosphere. This result suggests we should not expect planets resembling Neptune to retain substantial atmospheres inward of 2 days, although the exact boundary will depend on the atmospheric temperatures and composition. Meanwhile for the hot Jupiter, the ``New'' loss rate lags well behind the ``Evap'' rate for $P \gtrsim$ 0.5 days. 

Without considering the hydrodynamic details of the atmospheric escape, the mass transfer rate may also be set by the strength of the stellar tidal torque. During overflow, the escaping atmosphere can carry significant orbital angular momentum with it and form a circumstellar disk. Torques between the planet and disk can act in the opposite direction as the torque from the tide raised on the star, driving the remaining disk gas onto the star and returning its angular momentum to the planet \citep{1979MNRAS.186..799L}. Since overfilling the Roche lobe significantly increases the planet's mass loss rate, overfilling can also increase the disk mass and the strength of the outward torque. At the same time, mass loss would choke off if the disk pushed the planet outward of its Roche limit. Thus, in the limit that the gas disk quickly returns all of its angular momentum to the remaining planet, the mass loss rate may be set by the balance between the disk's outward torque and the stellar tide's inward torque \citep{1988ApJ...332..193V}. Assuming totally conservative mass transfer, the following expression for mass loss obtains \citep{2015ApJ...813..101V}:
\begin{equation}
\dot{M}_{\rm p, tidal} = \left(\dfrac{\dot{a}_{\rm tidal}}{2 a}\right) M_{\rm p},
\label{eqn:Mdot_tidal}
\end{equation}
where $\dot{a}_{\rm tidal}$ represents the change in semi-major axis due to the tide raised on the star \citep{2008ApJ...678.1396J}. Figure \ref{fig:Mdot_map} shows this ``Tidal'' mass loss rate for each hypothetical planet as black lines, assuming a tidal dissipation parameter for the star $Q_\star = 10^7$ \citep{2012ApJ...751...96P}. Traditionally, this loss model is only applied when the planet completely fills its Roche limit (when each planet reaches the appropriate vertical line in Figure \ref{fig:Mdot_map}) and is zero elsewhere. For visibility in the figure, we depict this rate as horizontal, black lines. For the hot Jupiter, the ``Tidal'' and ``New'' models are not too different as the planet encounters its Roche limit, but for the Neptune, the former rate is much smaller (owing the Neptune's small mass) than the latter at the Roche limit. This result suggests the ``New'' model overestimates the transfer rate if there is a balance between torques (in fact, they may not balance -- see Section \ref{sec:conclusions}).

\subsection{Candidate and Confirmed Planetary Systems}
\begin{figure}
\includegraphics[width=\textwidth]{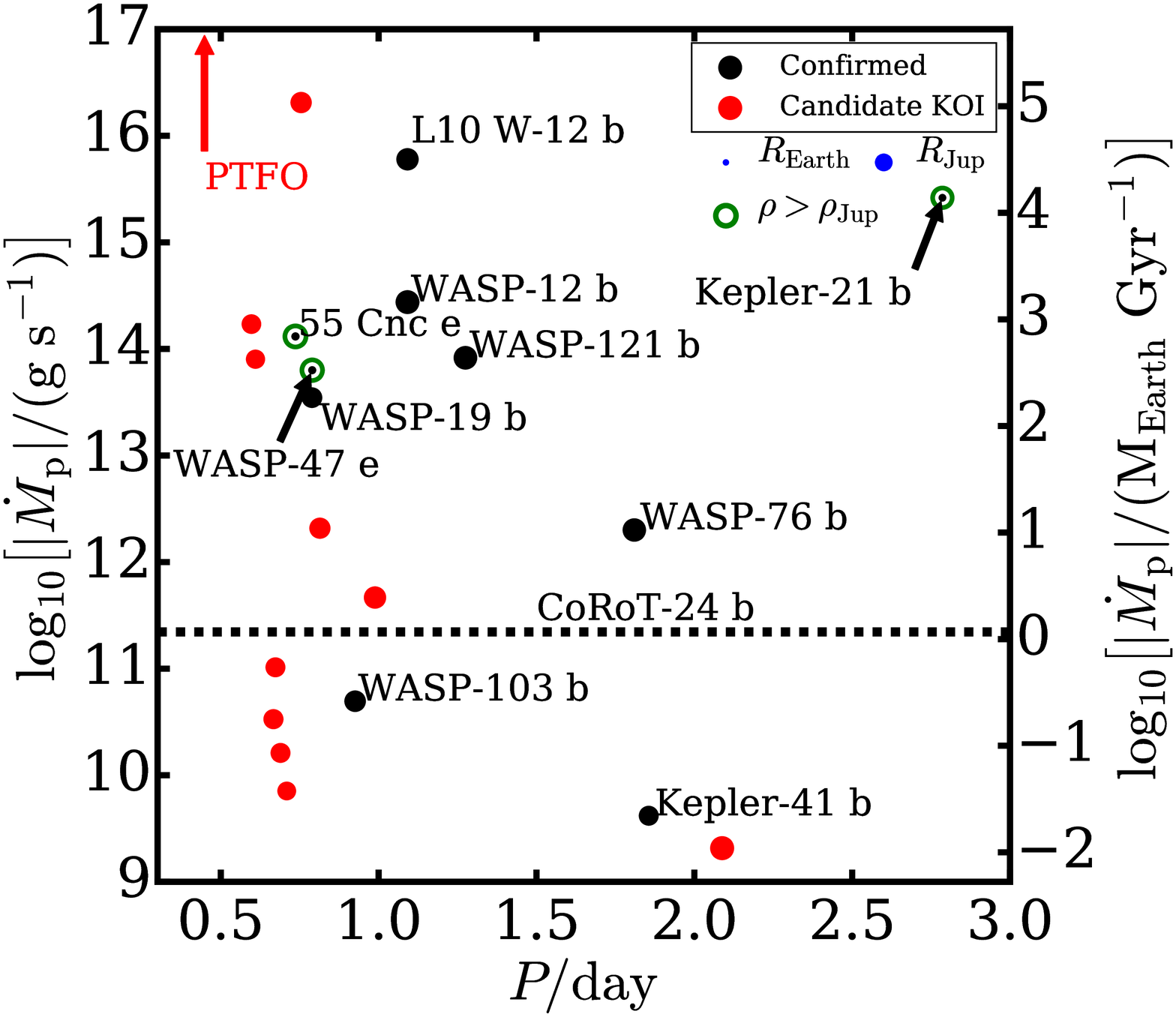}
\caption{Mass loss rates $\dot{M}_{\rm p}$ for confirmed (black circles) and candidate planets (red circles). Circle radius scales with object radius, as shown by the blue circles in the upper right. The figure shows our estimate for WASP-12 b and the loss estimated for WASP-12 b in the work of \citet{2010Natur.463.1054L} (labeled ``L10 W-12 b''). The loss rate for CoRoT-24 b is shown as a flat line since that planet's $P = 5.1$ days. The loss rate for PTFO8-8695 exceeds $10^{17}\ {\rm g\ s^{-1}} $ and is labeled with ``PTFO''. Green circles show planets whose bulk densities $\rho$ exceed Jupiter's $\rho_{\rm Jup}$ and may therefore not have a gaseous envelope to lose.}
\label{fig:Mdots_real}
\end{figure}

Figure \ref{fig:Mdots_real} shows our estimated mass loss rates for candidate and confirmed planetary systems. We consider planets and candidates with $P < 3$ days (with one exception), orbital eccentricity $e < 0.1$, and $1.6\ R_{\rm Earth} \le R_{\rm p} < 2\ R_{\rm Jup}$, large enough that they may have a significant gaseous envelope \citep{2015ApJ...801...41R} but small enough that they are likely planets. The boundary between planets that are bare rock and those with a non-negligible gaseous envelope is still the subject of active research. In Figure \ref{fig:Mdots_real}, we indicate those planets with bulk densities $\rho$ greater than Jupiter's bulk density $\rho_{\rm Jup} $ with green rings. They may or may not have gaseous envelopes. The data for all but one of the objects in the figure were retrieved on \catalogretrievaldate\ from the \anchor{http://exoplanetarchive.ipac.caltech.edu/}{NASA Exoplanet Archive} served up by the \anchor{http://archive.stsci.edu/}{MAST service}. The data for PTFO8-8695 come from \citet{2013ApJ...774...53B}. With sizes that scale linearly with the object's radius, black circles represent confirmed planets with mass and radius estimates, and red circles represent candidate Kepler Objects of Interest (KOIs). Table 2 provides the system parameters and results in tabular form. Since KOI candidates do not have known masses, we omit density estimates for them.

For candidate planets, mass estimates are unavailable, so for those with $1.6\ R_{\rm Earth} \le R_{\rm p} \le 4\ R_{\rm Earth}$, we used the density relationship from \citet{2014ApJ...783L...6W} to convert $R_{\rm p}$ to $M_{\rm p}$, although this conversion involves considerable uncertainty since, in this size range, $R_{\rm p}$ is much more sensitive to the fractional mass in the atmosphere than to the total $M_{\rm p}$ \citep{2014ApJ...792....1L}. For larger planets, we assumed $M_{\rm p} = 1\ M_{\rm Jup}$ since, for Jupiter-like planets, $R_{\rm p}$ is insensitive to $M_{\rm p}$ \citep{1984plin.book.....H}. Since the host star masses are unavailable for the candidate systems, we also estimated $a$ from $P$, $R_*$, and $\log_{\rm 10}\left( g \right)$ by calculating the corresponding $M_*$ and applying Kepler's law. 

Our mass loss estimates span a very broad range. For example, CoRoT-24 b (labeled as ``C-24 b''), has $R_{\rm p} = 3.7\ R_{\rm Earth}$ and $M_{\rm p} < 5.7\ M_{\rm Earth}$, and $T_{\rm p} = 1070 \pm 140$ K \citep{2014A&A...567A.112A}. Assuming it has a gaseous envelope of H$_2$, the planet has an enormous atmospheric scale height of about 1100 km, nearly 10\% $R_{\rm p}$, so either the planet is rapidly shedding mass, its atmospheric mean molecular mass is much larger than that of H$_2$, or its physical and orbital parameters require updating. Indeed, finding implausibly large evaporative mass loss, \citet{2016arXiv160503595L} argue similarly.

We also highlight WASP-12 b \citep{2009ApJ...693.1920H}, labeled ``W-12b'' in Figure \ref{fig:Mdots_real}. We estimate a mass loss of about $3\times10^{14}$ g ${\rm s^{-1}}$, meaning the planet might lose its entire atmosphere (1.41 $M_{\rm Jup}$) lin about 300 Myrs. By contrast, \citet{2010Natur.463.1054L} report a loss rate for WASP-12 b of nearly 10$^{16}$ g ${\rm s^{-1}}$, giving an even shorter atmospheric lifetime of about 10 Myrs. This discrepancy arises because some of the system's parameters have been revised since \citet{2010Natur.463.1054L} was published and because we estimate a slightly different escape potential. 

We also estimate the mass loss rate for PTFO8-8695, a putative hot Jupiter with a 10-hour orbital period around a 2-Myr old T-Tauri star \citep{2012ApJ...755...42V}. Based on attributing variations in the object's transit light curve to spin-orbit precession, \citet{2013ApJ...774...53B} estimated a mass and radius for the object of 3 ${\rm M_{Jup}}$ and 1.64 ${\rm R_{Jup}}$, respectively (assuming $M_\star = 0.34\ {\rm M_{{\rm Sun}}}$). As in \citet{2010Natur.463.1054L}, if we use this inferred transit radius and account for the tidal distortion of the photosphere, we find that the planet may actually overfill its Roche lobe by more than 10 scale heights, although the uncertainties on the inferred radius allow for an underfilled lobe. Assuming the object just fills its Roche lobe, our new model returns a loss rate of $> 10^{20}$ g ${\rm s^{-1}}$, meaning the object, if a hot Jupiter, would lose its atmosphere in 2,000 years. If we apply the ``Tidal'' loss rate and assume a conservative $Q_\star = 10^7$ \citep{2016arXiv160600848P}, we still find a loss rate in excess of $10^{17}$ g ${\rm s^{-1}}$, meaning the atmosphere escapes in less than 2 Myr. Given that the system is only about 2 Myrs, perhaps this rate is reasonable. In any case, the object, if a planet, is probably shedding considerable mass, as suggested in \citet{2016arXiv160602701J}. However, more recent follow-up observations have suggested the signal originally attributed to a planet is more likely to be starspots, eclipses by circumstellar dust, or occultations of an accretion hotspot \citep{2015ApJ...812...48Y}.

Figure \ref{fig:Mdots_real} also shows some candidate KOIs might have very large mass loss rates. In some cases, this result is likely an overestimate arising from our simplified estimates of the planets' unknown masses and radii, or the objects are false positives. Indeed, \citet{2012A&A...545A..76S} found that the false positive rate among giant planet candidates with short periods is larger than for other types of candidate. The large mass loss rates estimated here may help tease out which are false positives.

\section{Discussion and Conclusions}
\label{sec:conclusions}

The new model presented here updates the Roche-lobe overflow model from \citet{1988A&A...202...93R}. We incorporate many of the corrections provided by that model, in particular accounting for the extended atmospheres of stars and gaseous planets. Where appropriate, our model agrees with that of \citet{1988A&A...202...93R}. In spite of its utility, however, the latter model has limited accuracy and applies only over a fairly narrow range for the donor-accretor mass ratio $0.05 \le q \le 25$. Our model applies over many orders of magnitude for $q$, making it suitable for stars and short-period gaseous planets alike. 

Mapping out mass loss rates predicted by our model over a range of planetary masses and orbits suggests significant loss rates for some hot Jupiter candidates. Hot Neptunes may experience significant loss out to relatively long-period orbits, owing to their lower surface gravities. Loss rates from our model for hot Neptunes drop below loss rates from evaporative mass loss outward of $P \sim 1$ days, at least for the fiducial evaporative loss rates estimated here. However, Roche-lobe overflow and evaporative mass loss are not two distinct processes -- they both involve an unbound, hydrodynamic outflow of gas, but in the former case, the donor's photosphere coincides very closely with the Roche-lobe, while in the latter, it does not. Orbital (or internal) evolution may gradually drive a donor initially experiencing evaporative mass loss into Roche-lobe overflow, and the transition between Roche-lobe overflow and evaporative mass loss probably plays an important role in sculpting the atmospheres of short-period planets, especially for hot Neptunes. 

\begin{figure}
\includegraphics[width=\textwidth]{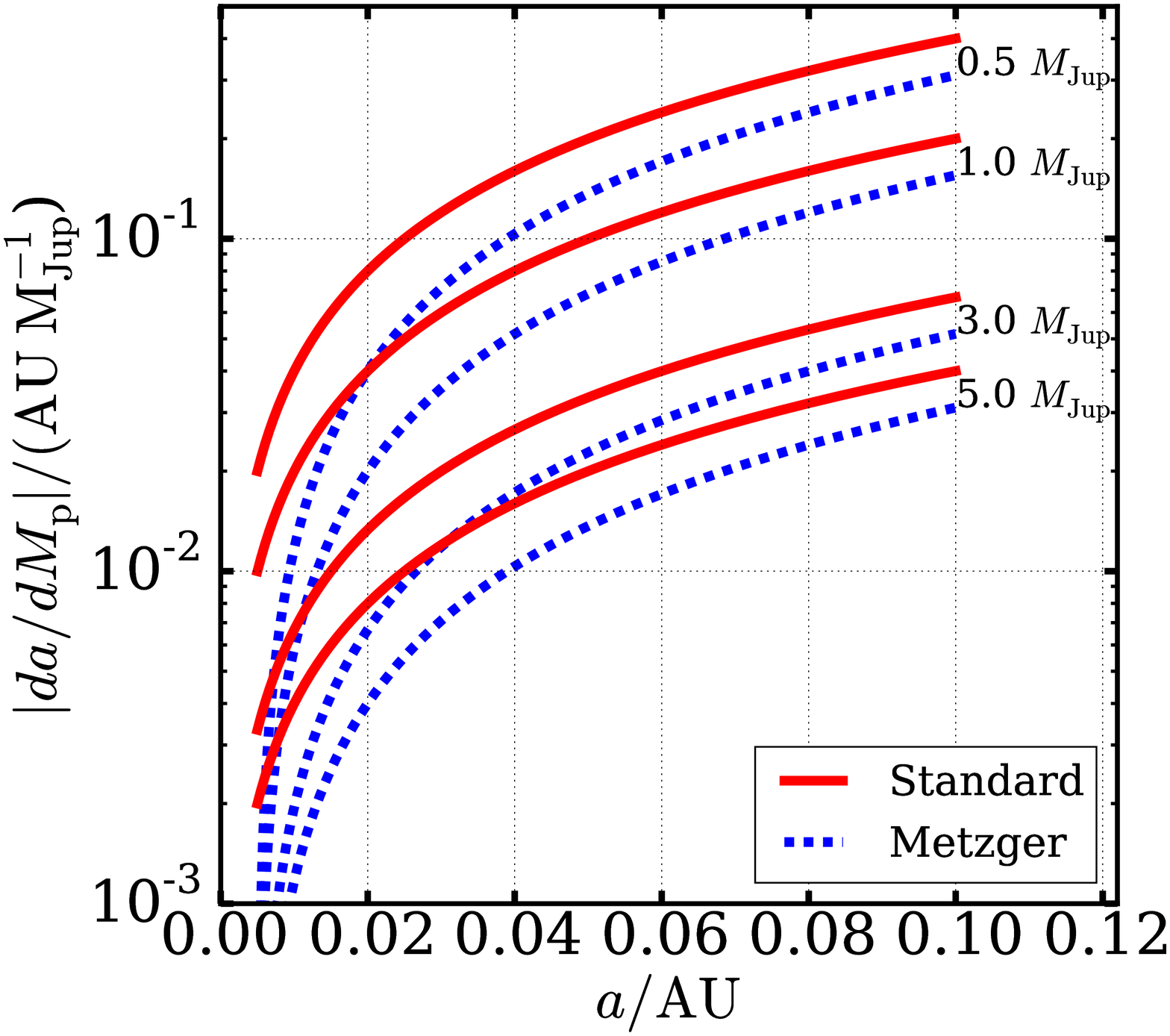}
\caption{Rate of change in semi-major axis $a$ for overflowing planets. The red, solid lines show the ``Standard'' expression for $da/dM_{\rm p}$, which assumes all angular momentum from the accretion disk is returned to the planet. The blue, dashed lines show the expression from \citet{2012MNRAS.425.2778M}, which accounts for loss of angular momentum from the innermost region of the accretion disk. Each pair of lines is labeled with the assumed $M_{\rm p}$-value.}
\label{fig:dadMp_vs_a}
\end{figure}

The orbital evolution that accompanies planetary Roche-lobe overflow also requires further investigation. One standard assumption is that orbital angular momentum carried away by the escaping gas is quickly returned to the donor object via tidal torques from the accretion disk \citep[e.g.][]{2016arXiv160300392J}, giving $da/dM_{\rm p} \approx -2 a / M_{\rm p}$ for $q \ll 1$. However, as explored in \citet{2015ApJ...813..101V}, the escaping gas may be quickly driven off without returning its angular momentum. As a consequence, the mass loss may enter a positive feedback: loss of mass increases the degree of Roche-lobe overfill, increasing the loss rate and driving more mass loss. Even if the gas is not quickly driven off, \citet{2012MNRAS.425.2778M} pointed out that not all of that angular momentum will be returned since the portion carried by gas striking the star is transmitted to the stellar rotation, giving instead $da/dM_{\rm p} = -2 \left( a / M_{\rm p} \right) \left( 1 - \sqrt{R_*/a} \right)$ for $q \ll 1$. Figure \ref{fig:dadMp_vs_a} compares the two expressions for $da/dM_{\rm p}$ over a range of $M_{\rm p}$ and $a$. For the shortest-period orbits, the ``Metzger'' expression drops rapidly compared to the ``Standard'' expression, meaning the loss of orbital angular momentum is exacerbated for the short-period planets. These effects potentially doom planets to unstable mass loss, possibly causing a planet to fall apart on dynamical timescales. 

Whether the mass loss actually becomes unstable, however, depends on how the rate of increase in mass loss compares to the rate at which the accretion disk transfers angular momentum, i.e. the disk's viscous timescale: if the disk very quickly returns the angular momentum (assuming such a disk forms), then it can accommodate an increased mass loss rate and the loss may remain stable. Indeed, one purpose for developing the mass loss model in \citet{1988A&A...202...93R} was to estimate the timescale for development of a mass loss instability independent of assumptions regarding the exchange of angular momentum. 

\citet{1988A&A...202...93R} argued that the timescale for a change in the mass loss rate $\tau_{\dot{M}}$ is given by 
\begin{equation}
\tau_{\dot{M}} = \dfrac{\dot{M}}{\ddot{M}} = \dfrac{H_{\rm p}}{R} \left\{ \dfrac{M}{\dot{M}} \left( \zeta_{\rm s} - \zeta_{\rm R} \right) + \dfrac{1}{\tau_{R}} - \dfrac{2}{\tau_{\rm J}} \right\}^{-1}, 
\label{eqn:Ritters_instability_timescale}
\end{equation}
where $M/R$ represents the mass/photospheric radius of the donor ($R_{\rm p}$ for planets); $H_{\rm p} = H_{\rm p, 0}/\gamma$ is the donor's modified atmospheric scale height (Equation \ref{eqn:modified_scale_height}); $\zeta_{\rm s} \equiv \left( \partial \ln R/\partial \ln M \right)$ at constant entropy (s); $\zeta_{\rm R} \equiv \left(\partial\ln r_{\rm R}/\partial \ln M \right)$, i.e. the logarithmic change in the effective Roche-lobe radius with mass (which is $\approx 1/3$ for planets); $\tau_{\rm R} \equiv \partial \ln R/\partial t$, i.e. the change in radius at constant mass; and $\tau_{\rm J}$ is the timescale over which orbital angular momentum changes. We can adapt this expression for hot Jupiters and Neptunes to gauge whether they are prone to unstable mass loss. 

The equation below shows the instability timescales for typical hot Jupiters and Neptunes:
\begin{eqnarray}
\textrm{Jupiters:}\ \tau_{\rm \dot{M}_2} = \left( \dfrac{\rm 1000\ km}{\rm 70000\ km} \right) \left\{ \dfrac{10^{18}\ {\rm g\ s^{-1}}}{2\times10^{30}\ {\rm g}} \left(0 - \frac{1}{3} \right) + \dfrac{1}{\rm 10\ Gyr} - \dfrac{1}{1\ {\rm Gyr}}\right\}^{-1} \sim 3000\ {\rm yr} \nonumber \\
\textrm{Neptunes:}\ \tau_{\rm \dot{M}_2} = \left( \dfrac{\rm 2000\ km}{\rm 24000\ km} \right) \left\{ \dfrac{10^{17}\ {\rm g\ s^{-1}}}{10^{29}\ {\rm g}} \left(0 - \frac{1}{3} \right) + \dfrac{1}{\rm 10\ Gyr} - \dfrac{1}{10\ {\rm Gyr}}\right\}^{-1} \sim 8000\ {\rm yr}. \nonumber
\end{eqnarray}
As suggested in the results from \citet{2007ApJ...659.1661F}, we have assumed a constant radius with mass $\zeta_{\rm s} \sim 0$, a very slow contraction of the planet's radius \citep{2006ApJ...650..394A}, and estimated the timescale for orbital angular momentum change using the tidal model in \citet{2016arXiv160300392J} and assuming $a \approx$ 0.01 AU. 

The viscous timescale for the resulting accretion disk $\tau_{\rm visc} \sim a^2/\left( \alpha v_{\rm th} H \right) = a \Omega/\left( \alpha v_{\rm th}^2\right)$ \citep{1981ARA&A..19..137P}, where $\Omega$ is the orbital mean motion, $H$ the disk scale height ($= v_{\rm th}/\Omega$), $\alpha$ the viscosity parameter $\sim 0.01$ \citep{2009ApJ...705.1206C}, and $v_{\rm th}$ the speed of sound $\sim$ 5 km ${\rm s^{-1}}$ \citep{2010Natur.463.1054L}. These parameter choices suggest $\tau_{\rm visc} \lesssim$ 10 yr, much less than the instability timescale for either planet. Detailed hydrodynamic modeling is required to robustly investigate whether such mass loss is actually stable, but these simple scaling relations suggest it should be. 

This result has important implications for the origins of ultra-short-period planets. \citet{2015ApJ...813..101V} showed that stable Roche-lobe overflow drives the overflowing planets out to orbital periods of several days, and \citet{2016arXiv160300392J} argued subsequent tidal decay is unlikely to bring most of them back to $P <$ 1 day. Thus, stable Roche-lobe overflow means ultra-short-period planets are probably not the remnants of disrupted gaseous planets.

Our model suggests many confirmed and candidate short-period planets may be experiencing significant mass loss. In particular, even with $P = 5.1$ days, CoRoT-24 b's low surface gravity and high effective temperature mean it could be losing mass at a prodigious rate. On the other hand, WASP-12 b may not be losing mass as quickly as was recently estimated \citep{2010Natur.463.1054L}. Follow-up observations of some of \kepler's short-period candidate planets may reveal mass loss signatures, perhaps in the form of warm CO emission from the resulting accretion disk, as suggested by \citet{2010Natur.463.1054L}. If a gaseous planet, PTFO8-8695 is rapidly losing its atmosphere.

Our model and the mass estimates we have presented here involve several important assumptions and approximations upon which future work could improve. In particular, the isothermal approximation represents an important limitation, and future work should explore if improved approximations of radiative and adiabatic heating and cooling can be incorporated efficiently enough for inclusion alongside models coupling mass loss and thermal and orbital evolution. In principle, there is also some interplay between the evaporative mass loss and the Roche-lobe overflow, which presumably modifies the mass loss rates and orbital evolution discussed here. 

\begin{figure}
\includegraphics[width=\textwidth]{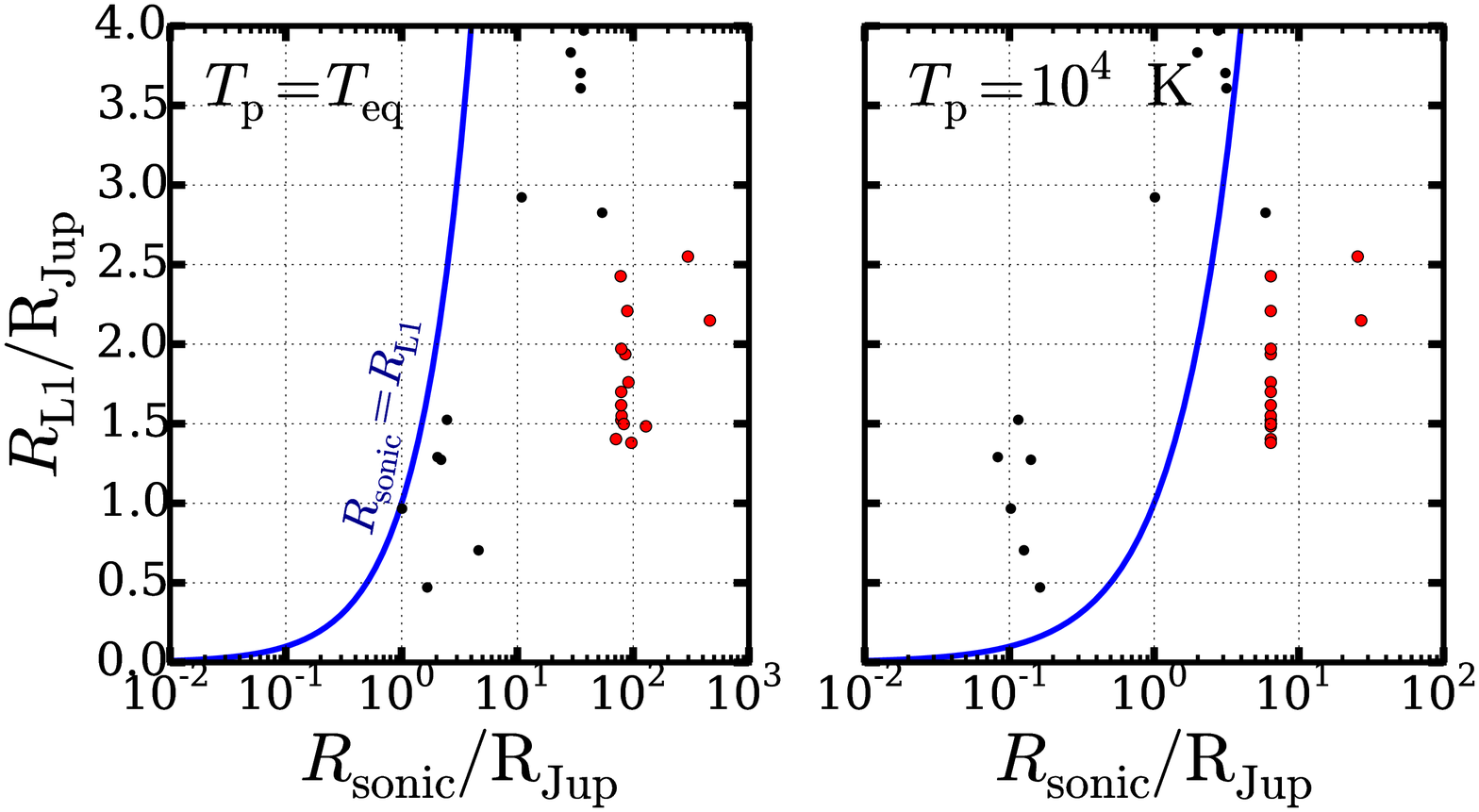}
\caption{Distance from a planetary object's center to the Lagrange point L1, $R_{\rm L1}$ compared to the sonic radius $R_{\rm sonic} = G M_{\rm p}/2 v_{\rm th}^2$. The left panel provides this comparison, assuming the atmosphere has the radiative equilibrium temperature, while the right panel assumes an atmospheric temperature of $10^4$ K. As in Figure \ref{fig:Mdots_real}, the black dots indicate confirmed planets, the red dots indicate candidate planets. The blue line shows $R_{\rm sonic} = R_{\rm L1}$. }
\label{fig:comparing_sonic_and_Roche_radii}
\end{figure}

Indeed, many of the planets depicted in Section \ref{sec:results} may lie near the crossover between evaporative mass loss and Roche-lobe overflow. We can qualitatively explore that crossover by roughly estimating the radius $R_{\rm sonic} = G M_{\rm p}/2 v_{\rm th}^2$ at which we would expect an outflow to become transonic, i.e. crosses over from sub- to supersonic, and then comparing $R_{\rm sonic}$ to the radius of the L1 Lagrange point $R_{\rm L1}$ (as measured along the x-axis connecting the planet and star). Figure \ref{fig:comparing_sonic_and_Roche_radii} shows these two radii for each object depicted in Figure \ref{fig:Mdots_real}. The left panel assumes the same atmospheric temperatures as in Figure \ref{fig:Mdots_real}, while the right panel assumes $T_{\rm p} = 10^4$ K, which may be more appropriate for some objects \citep{2011ApJ...728..152T}.

As shown in the left panel, the lower atmospheric temperatures give $R_{\rm L1} < R_{\rm sonic}$ for the vast majority of systems, meaning, if atmospheres are escaping at all, it is via Roche-lobe overflow. By contrast, the right panel assumes $T_{\rm p} = 10^4$ K, which is characteristic of the thermosphere of close-in planets, where EUV heating is balanced by atomic line cooling or adiabatic cooling in a planetary wind \citep{2004Icar..170..167Y}. Nearly all systems would be in the Roche-lobe overflow limit using the equilibrium temperature, but in the limit that the gas is at the thermosphere temperature, a significant fraction of systems have $R_{\rm L1} > R_{\rm sonic}$. Their mass loss rates might more accurately be described by the photoevaporation model. Thus, exploration of the regime between evaporative mass loss and Roche-lobe overflow may be critical for understanding many of the shortest-period planets. 

\acknowledgments
The authors acknowledge helpful input from Tiffany Kataria. Thorough reviews from a referee also greatly improved the paper, for which we are grateful.

Using Foreman-Mackey's \anchor{http://dan.iel.fm/kplr/}{kplr} python package, this research accessed the Mikulski Archive for Space Telescopes (MAST). STScI is operated by the Association of Universities for Research in Astronomy, Inc., under NASA contract NAS5-26555. Support for MAST for non-HST data is provided by the NASA Office of Space Science via grant NNX13AC07G and by other grants and contracts. This study used data collected by the \kepler\ mission, funding for which is provided by the NASA Science Mission directorate. The figures, data, and \LaTeX\ manuscript for this study are available at \url{http://www.astrojack.com/research}. This study was supported by NASA Grant NNX13AQ62G.

{\it Facilities:} \facility{\kepler}

\appendix
\section{Deriving the Volume-Averaged Potential}
\label{apx:potential}
Define $r_{\rm \Phi}(\theta, \phi)$ to be the equipotential surface for a given $\Phi$-value [note that $r_{\rm \Phi}$ is not the same as $\rv$]. The volume around the donor enclosed by that equipotential surface is
\be
V(\Phi) & = & \int_{V}  dV = \int d\Omega \int_0^{r_\phi(\theta,\phi)} dr r^2 =  \frac{1}{3} \int d\Omega\ r_\Phi^3(\theta,\phi) \equiv \frac{4\pi}{3} \rv^3(\Phi),
\label{eq:VPhi}
\ee
where $d\Omega$ is the solid angle. Brackets $\langle \rangle$ denote an average over solid angle, so that
$r_{\rm v}^3 = \langle r_\Phi^3 \rangle$. The equipotential volume $V(\Phi)$ can straightforwardly be computed by numerical integration using the first form in Equation \ref{eq:VPhi}, by only including points if their potential is less than $\Phi$. Numerical integration has the advantage that the full Roche potential in Equation \ref{eq:phi_full} can be used, with no approximation, but the disadvantage is that it is time-consuming for the integrals to be computed accurately. A much faster method, appropriate for use in stellar evolution codes, is to use analytic approximations. 

First, we define a modified potential $\psi$ as 
\be
\psi & \equiv & -  \left(  \Phi  +  \frac{GM_{\rm a}}{a} \left[ 1 + \left( \frac{r}{a} \right) \cos \theta \right] + \frac{1}{2} \frac{GM_{\rm a}^2}{a(M_{\rm d}+M_{\rm a})} \right),
\label{eqn:psi_potential}
\ee
i.e., the negative of the gravitational potential with constant terms and terms whose volume average is zero subtracted out. Taking the ratio $r/a$ to be small (i.e., considering the potential only near the donor), we can expand $\psi$ as
\be
\psi & = & \frac{G M_{\rm d}}{r} \nonumber \\ 
& + & \frac{G M_{\rm a}}{a} \left[ \frac{1}{2} \left( 3 \cos^2 \theta - 1 \right) \left( \frac{r}{a} \right)^2 + \frac{1}{2} \left( 5 \cos^3 \theta - 3 \cos \theta \right) \left( \frac{r}{a} \right)^3 \right] \nonumber \\
& + & \frac{1}{2} \frac{G \left( M_{\rm d} + M_{\rm a} \right)}{a^3} \left[ \left( r^2 \cos^2 \theta - 2 \left( \frac{M_{\rm a}}{M_{\rm d} + M_{\rm a}} \right) a r \cos \theta + \left( \frac{M_{\rm a}}{M_{\rm d} + M_{\rm a}} \right)^2 a^2 \right) + r^2 \sin^2 \theta \cos^2 \phi \right]
\label{eqn:psi_initial_expansion}
\ee

First, define a zeroth-order radius
\be
r_0(\psi) = \frac{G M_{\rm d}}{\psi}.
\ee
That is, at the zeroth order, $\psi \approx G M_{\rm d}/r_0$. Of course, the actual potential is not constant along the sphere with radius $r_0$. However, since $\psi$ is dominated by the donor's gravity inside of its Roche lobe, we can write the radius for that potential contour $r_\Phi$ in terms of an expansion in powers of $r_0(\psi)/a$:
\be
r_{\rm \Phi} & =& r_0 + r_1 + r_2,
\ee
where each term will turn out to be smaller than the previous one by a power of $(r_0/a)^3$. Plugging this expression into \ref{eqn:psi_initial_expansion} and expanding $r_{\rm \Phi}$, we can match up orders on the left- and right-hand of the resulting equation. For example, at the zeroth order 
\be
\psi \approx \frac{G M_{\rm d}}{r_0}.
\label{eqn:r1}
\ee
At the first order,
\be
0 \approx &-&\frac{G M_{\rm d}}{r_0^2} r_1 \nonumber \\
& + & \frac{G M_{\rm a}}{a^3} \left( \frac{3}{2} \cos^2 \theta - \frac{1}{2} \right) r_0^2 \nonumber \\
& + & \frac{1}{2} \frac{G \left( M_{\rm a} + M_{\rm d} \right)}{a^3} \left( \cos^2 \theta + \sin^2 \theta \cos^2 \phi \right) r_0^2, 
\label{eqn:r1}
\ee
giving 
\be
\left( \frac{r_1}{r_0} \right) = \left[ \left( \frac{M_{\rm a}}{M_{\rm d}} \right) \left( \frac{3}{2} \cos^2 \theta - \frac{1}{2} \right) + \frac{1}{2} \left( \frac{M_{\rm d} + M_{\rm a}}{M_{\rm d}} \right) \left( \cos^2 \theta + \sin^2 \theta \cos^2 \phi \right) \right] \left( \frac{r_0}{a} \right)^3.
\ee
The next order gives 
\be
\left( \frac{r_2}{r_0} \right) = 3 \left( \frac{r_1}{r_0} \right)^2.
\label{eqn:r2}
\ee
We can average these expressions over the unit sphere to calculate the volume of the potential contour, as in Equation \ref{eq:VPhi}:
\begin{eqnarray}
r_{\rm v}^3 & \approx & r_0^3 + 3r_0^2 \langle r_1 \rangle + 3r_0^2 \langle r_2 \rangle + 3r_0 \langle r_1^2 \rangle \nonumber \\
& = & r_0^3 \left[ 1 + \left( \frac{r_0}{a} \right)^3 \left( \frac{ M_{\rm d}+M_{\rm a} }{ M_{\rm d} } \right) 
+ \frac{4}{5} \left( \frac{r_0}{a} \right)^6 \left\{ \frac{ 2(M_{\rm d}+M_{\rm a} )^2 + M_{\rm a} (M_{\rm d}+M_{\rm a} )
+ 3 M_{\rm a}^2}{M_{\rm d}^2} \right\} \right].
\label{eq:rv_of_psi}
\end{eqnarray}

Equation \ref{eq:rv_of_psi} expresses the volume in terms of the potential. It remains to invert this expression, to express the potential as a function of the volume. To accomplish this, define the series expansion
\be
\psi(r_{\rm v}) = \psi_0 + \psi_1 + \psi_2
\label{eq:psi_expansion}
\ee
where each successive term is smaller by a factor $(r_{\rm v}/a)^3$. At the zeroth order, 
\be
r_{\rm v} \approx r_0 = \frac{G M_{\rm d}}{\psi_0}.
\ee

Expressing $r_{\rm v}^3$ in terms of this expansion gives 
\begin{eqnarray}
r_{\rm v}^3 \approx r_0^3 & = & \left( \frac{G M_{\rm d}}{\psi_0 + \psi_1 + \psi_2} \right)^3 \nonumber \\
	& \approx & \left( \frac{G M_{\rm d}}{\psi_0} \right)^3 \left[ 1 - 3 \left( \frac{\psi_1}{\psi_0} \right) - 3 \left( \frac{\psi_2}{\psi_0} \right) - 6 \left( \frac{\psi_1}{\psi_0} \right)^2 \right] \nonumber \\
	& \approx & r_{\rm v}^3 \left\{ 1 - 3 \left( \frac{\psi_1}{\psi_0} \right) - 3 \left( \frac{\psi_2}{\psi_0} \right) - 6 \left( \frac{\psi_1}{\psi_0} \right)^2 \right\} \nonumber \\ 
	& + &  r_{\rm v}^3 \left\{ \left( \frac{r_v}{a} \right)^3 \left( \frac{ M_{\rm d}+M_{\rm a} }{ M_{\rm d} } \right) + \frac{4}{5} \left( \frac{r_v}{a} \right)^6 \left[ \frac{ 2( M_{\rm d}+M_{\rm a} )^2 + M_{\rm a} (M_{\rm d}+M_{\rm a} )+ 3 M_{\rm a}^2}{M_{\rm d}^2} \right] \right\}
\label{eq:rv3_in_terms_of_potential_expansion}
\end{eqnarray}

Again, matching up terms of the same order gives
\be
\frac{\psi_1}{\psi_0} & = & \frac{1}{3} \left( \frac{ M_{\rm d}+M_{\rm a} }{ M_{\rm d} } \right) \left( \frac{r_{\rm v}}{a} \right)^3
\\
\frac{\psi_2}{\psi_0} & = & - 4 \left( \frac{\psi_1}{\psi_0}  \right)^2 + \frac{4}{15} \left( \frac{r_{\rm v}}{a} \right)^6
 \left( \frac{ 2(M_{\rm d}+M_{\rm a} )^2 + M_{\rm a} (M_{\rm d}+M_{\rm a} )
+ 3 M_{\rm a}^2}{M_{\rm d}^2} \right).
\ee
Combining these results gives the final expression
\be
\Phi(\rv) = &-& \left(  \frac{GM_{\rm a}}{a} + \frac{ GM_{\rm a}^2 }{2a(M_{\rm d}+M_{\rm a})} \right)
\nonumber \\ & - & 
 \frac{GM_{\rm d}}{\rv} \left[ 1 + \frac{1}{3} \left( \frac{M_{\rm d}+M_{\rm a}}{M_{\rm d}} \right) \left( \frac{\rv}{a} \right)^3
 \right. \nonumber \\ & + & \left.
 \frac{4}{45} \left( \frac{(M_{\rm d}+M_{\rm a})^2 + 9M_{\rm a}^2 + 3M_{\rm a}(M_{\rm d}+M_{\rm a})}{M_{\rm d}^2}\right) \left( \frac{\rv}{a} \right)^6
\right].
\label{eq:appendix_phi_vs_rv}
\ee

\bibliography{Jacksonetal_2015_RLO}
\bibliographystyle{apj}

\input{Jacksonetal_2016_RLO_pl_params_table.tex}

\end{document}

%% file: Ritter_Table_A1.tex
\begin{deluxetable}{ccccc}




\tablecaption{Reproduction of a Portion of Table A1 from \citet{1988A&A...202...93R}.\label{tbl:ritter_table_a1}}

\tablenum{1}

\tablehead{\colhead{$M_{\rm d}$} & \colhead{$r_{\rm ph}$} & \colhead{$T_{\rm eff}$} & \colhead{$\mu$} & \colhead{$\rho_{\rm ph}$} \\ 
\colhead{($M_{\rm Sun}$)} & \colhead{($R_{\rm Sun}$)} & \colhead{(K)} & \colhead{molar mass} & \colhead{(g cm$^{-3}$)} } 

\startdata
1.2 & 1.17 & 6480 & 1.31 & 2.5$\times10^{-7}$ \\
0.7 & 0.67 & 4430 & 1.31 & 1.3$\times10^{-6}$ \\
0.6 & 0.59 & 3900 & 1.26 & 1.4$\times10^{-6}$ \\
0.5 & 0.52 & 3520 & 1.33 & 2.0$\times10^{-6}$ \\
0.25 & 0.25 & 3410 & 1.31 & 1.6$\times10^{-5}$ \\
0.085 & 0.1 & 2740 & 1.33 & 5.0$\times10^{-5}$ \\
\enddata




\end{deluxetable}

%% file: Jacksonetal_2016_RLO_pl_params_table.tex



\begin{deluxetable}{ccccccccccc}

\tabletypesize{\tiny}




\tablecaption{Planetary System Parameters}

\tablenum{2}

\tablehead{\colhead{Object Name} & \colhead{$M_{\rm p}$} & \colhead{$R_{\rm p}$} & \colhead{$\rho$} & \colhead{Estimated $T_{\rm p}$} & \colhead{$P$} & \colhead{$a$} & \colhead{$M_{\rm s}$} & \colhead{$\dot{M}_{\rm p}$} & \colhead{$\dot{M}_{\rm p}$} & \colhead{Reference} \\ 
\colhead{} & \colhead{(${\rm M_{Jup}}$)} & \colhead{(${\rm R_{Jup}}$)} & \colhead{(${\rm g\ cm^{-3}}$)} & \colhead{(K)} & \colhead{(days)} & \colhead{(AU)} & \colhead{(${\rm M_{Sun}}$)} & \colhead{(${\rm g\ s^{-1}}$)} & \colhead{(${\rm M_{Earth}\ Gyr^{-1}}$)} & \colhead{} } 

\startdata
Kepler-21 b &  0.01598 &  0.146 &  6.40 &  2411 &  2.78578 &  0.04272 &  1.41 &  2.62e+15 &  1.38e+04 &  a \\
Kepler-41 b &  0.56 &  1.29 &  0.33 &  2126 &  1.8555582 &  0.03101 &  1.15 &  4.17e+09 &  2.21e-02 &  a \\
WASP-76 b &  0.92 &  1.83 &  0.20 &  2595 &  1.809886 &  0.03300 &  1.46 &  2.00e+12 &  1.06e+01 &  a \\
55 Cnc e &  0.02542 &  0.17 &  6.40 &  2325 &  0.736539 &  0.01544 &  0.91 &  1.31e+14 &  6.91e+02 &  a \\
WASP-12 b &  1.47 &  1.9 &  0.27 &  3071 &  1.0914203 &  0.02340 &  1.43 &  2.75e+14 &  1.45e+03 &  a \\
WASP-121 b &  1.183 &  1.865 &  0.24 &  2806 &  1.2749255 &  0.02544 &  1.35 &  8.26e+13 &  4.36e+02 &  a \\
WASP-19 b &  1.069 &  1.392 &  0.49 &  2497 &  0.78883899 &  0.01634 &  0.90 &  3.49e+13 &  1.84e+02 &  a \\
WASP-103 b &  1.49 &  1.528 &  0.55 &  2984 &  0.925542 &  0.01985 &  1.22 &  4.95e+10 &  2.62e-01 &  a \\
CoRoT-24 b &  0.018 &  0.33 &  0.90 &  1112 &  5.1134 &  0.05600 &  0.91 &  2.21e+11 &  1.17e+00 &  a \\
WASP-47 e &  0.02863 &  0.162 &  8.50 &  2618 &  0.789636 &  0.01730 &  1.11 &  6.32e+13 &  3.34e+02 &  a \\
K01169.01 &  1.0 &  1.29 &  ... &  1927 &  0.689209786 &  0.01546 &  1.04 &  1.62e+10 &  8.58e-02 &  a \\
K02607.01 &  1.0 &  1.49 &  ... &  1932 &  0.754458321 &  0.01612 &  0.98 &  2.05e+16 &  1.08e+05 &  a \\
K04325.01 &  1.0 &  1.19 &  ... &  2143 &  0.609926492 &  0.01442 &  1.08 &  8.01e+13 &  4.23e+02 &  a \\
K04595.01 &  1.0 &  1.25 &  ... &  1577 &  0.597017651 &  0.01231 &  0.70 &  1.71e+14 &  9.02e+02 &  a \\
K04685.01 &  1.0 &  1.65 &  ... &  1781 &  0.988493125 &  0.02008 &  1.11 &  4.65e+11 &  2.45e+00 &  a \\
K02480.01 &  1.0 &  1.31 &  ... &  1181 &  0.666826778 &  0.01226 &  0.55 &  3.36e+10 &  1.77e-01 &  a \\
K06262.01 &  1.0 &  1.29 &  ... &  1834 &  0.673141323 &  0.01481 &  0.96 &  1.03e+11 &  5.44e-01 &  a \\
K06532.01 &  1.0 &  1.47 &  ... &  1931 &  0.813998777 &  0.01675 &  0.95 &  2.08e+12 &  1.10e+01 &  a \\
K07559.01 &  1.0 &  1.13 &  ... &  2417 &  0.708965461 &  0.01643 &  1.18 &  7.12e+09 &  3.76e-02 &  a \\
K07594.01 &  1.0 &  1.93 &  ... &  2003 &  2.08802799 &  0.03517 &  1.33 &  2.07e+09 &  1.09e-02 &  a \\
PTFO8-8695 &  3.0 &  1.64 &  ... &  2268 &  0.44841 &  0.00800 &  0.34 &  2.14e+20 &  1.13e+09 &  b \\
\enddata



\tablerefs{a - NASA Exoplanet Archive, b - \citet{2013ApJ...774...53B}}

\end{deluxetable}